\documentclass[twocolumn,prb,amsmath,amssymb,floatfix,superscriptaddress]{revtex4-2} 
\usepackage{amsmath}
\usepackage{amssymb}
\usepackage{graphicx}
\usepackage{bm}
\usepackage{color}
\usepackage{hyperref}
\usepackage{graphicx}
\usepackage{upgreek}
\usepackage{scalerel}
\usepackage{multirow}
 
 \usepackage[bb=dsserif]{mathalpha}
 
\usepackage{pgffor}
\usepackage{pdfpages}
\usepackage{mathdots}
\usepackage{float}
\usepackage{silence}
\usepackage{physics} 
\usepackage{lscape}   
\usepackage{color, colortbl}
\usepackage[table]{xcolor}
\usepackage{comment}  
\makeatletter
\AtBeginDocument{\let\LS@rot\@undefined}
\makeatother

\begin{document}

\title{Andreev exceptional points in Josephson junctions formed by minimal Kitaev chains}
\author{Jorge Cayao}
\email[Corresponding author: ]{jorge.cayao@physics.uu.se}
\affiliation{Department of Physics and Astronomy, Uppsala University, Box 516, S-751 20 Uppsala, Sweden}
\date{\today}

\author{Masatoshi Sato}
\affiliation{Center for Gravitational Physics and Quantum Information, Yukawa Institute for Theoretical Physics, Kyoto University, Kyoto 606-8502, Japan}

\begin{abstract}
We consider Josephson junctions formed by two minimal Kitaev chains and investigate how the interplay between non-Hermiticity and superconducting phase difference enables the realization of stable topological states that do not exist in the Hermitian realm. In particular, we focus on non-Hermiticity produced by coupling the minimal Kitaev chain Josephson junction to normal reservoirs, which renders the system open and characterized by a complex Andreev spectrum. Interestingly, we find that this complex spectrum hosts second-order exceptional points, where a pair of eigenvalues and their respective eigenvectors coalesce, and are fully controlled by the superconducting phase difference. Depending on the spatial unequal distribution of non-Hermiticity, these Andreev exceptional points can appear at zero or finite energies connecting stable energy lines protected by non-Hermitian topology. Moreover, by tuning the system parameters, such as onsite energies, non-Hermiticity, or electron cotunneling, the Andreev exceptional points give rise to Andreev exceptional lines enclosing protected two-dimensional zero real energy areas.  We also discuss potential detection schemes of Andreev exceptional points by using local and nonlocal conductance signatures. Our results demonstrate the utility of non-Hermiticity from normal reservoirs as a useful resource for engineering non-Hermitian topological phases in minimal Kitaev chain Josephson junctions.
\end{abstract}
\maketitle

\section{Introduction}
Advances in superconductor-semiconductor hybrids have recently enabled the experimental realization of minimal Kitaev chains \cite{dvir2023realization,Haaf2024,Zatelli_2024,bordin2025enhanced,ten2025observation,bordin2026probing,van2026single,ryll-qb42,zatelli2026MZM}, where two coupled quantum dots (QDs) develop the necessary ingredients for realizing a two-site Kitaev chain \cite{Leijnse2012,Sau2012,PhysRevB.90.220501}. In particular, the QDs are coupled by electron cotunneling (ECT) and crossed Andreev reflections (CARs), which, respectively, characterize electron tunneling and the exotic $p$-wave nature associated with the underlying superconductivity of regular Kitaev chains \cite{tanaka2011symmetry,sato2016majorana,sato2017topological,Aguadoreview17,lutchyn2018majorana,Cayao2020Oddfrequency,flensberg2021engineered,tanaka2024theory}. It is also known that, at the sweet spot where CAR equals ECT, minimal Kitaev chains host fully nonlocal  Majorana zero modes \cite{Leijnse2012,PhysRevB.110.125408,PhysRevB.106.L201404,PhysRevResearch.5.043182,PRXQuantum.5.010323,liu2024enhancing,PhysRevB.110.245144,PhysRevB.109.035415,PhysRevB.110.165404,PhysRevB.110.245412,nitsch2024tetron,kotetes2024nonRecifourpi,kumar2026Awoga,PhysRevB.111.115419,vimal2025EntMKC,Seoane_Souto_2024,pitavidal2025qubit,w8f7-6cyy,1jqt-3ht5,bg4z-vx9m}, which, despite their bulk topological triviality, are symmetry-protected  \cite{hf7s-f7tj,cayaosatosymmetryMKC}.  While this sweet-spot symmetry protection  can be useful for certain tasks, it is not topologically protected.

An interesting way to achieve topological protection is by exploiting the inherent open nature of physical systems coupled to reservoirs \cite{datta1997electronic}, where dissipative effects can induce spectral degeneracies known as exceptional points (EPs) that define non-Hermitian topology \cite{PhysRevLett.86.787,PhysRevX.9.041015,PhysRevX.8.031079,doi:10.7566/JPSCP.30.011098,OS23,RevModPhys.93.015005}. EPs are special because at these topological points, both eigenvalues and eigenvectors coalesce \cite{el2018non,ozdemir2019parity,doi:10.1080/00018732.2021.1876991}, unlike Hermitian degeneracies where only eigenvalues merge. These non-Hermitian ideas have already been utilized in several trivial superconducting systems for realizing non-Hermitian topological matter, including conventional superconducting setups \cite{JorgeEPs,avila2019non,PhysRevLett.123.097701,PhysRevB.99.041202,PhysRevB.99.165145,PhysRevB.105.094502,PhysRevB.107.104515,PhysRevB.110.085414} and Josephson junctions \cite{cayao2023nonhermitian,li2023anomalous}; see also Refs.\,\cite{PhysRevB.108.L060506,shen2024nonhermitian,pino2024thermodynamics,Ohnmacht2024,cayao2024non,li2025EP,PhysRevB.111.064517,ogino2025,solow2025EP,JunjieNHDiode,cayaoSatoNH4MZMs,9jdy-b418,mhwj-5c85,gbhj-65b7,capecelatro20260pi,PhysRevB.103.235438}. However, only limited attention has been devoted to   designing of topological phases using  minimal Kitaev chains despite the evident potential \cite{cayaominimalkitaev,PhysRevB.109.L161404}. It is, for instance, largely unknown how the Josephson effect  in minimal Kitaev chains would induce and control the formation of EPs and, hence, non-Hermitian topology.

 \begin{figure}[!b]
\centering
\includegraphics[width=0.45\textwidth]{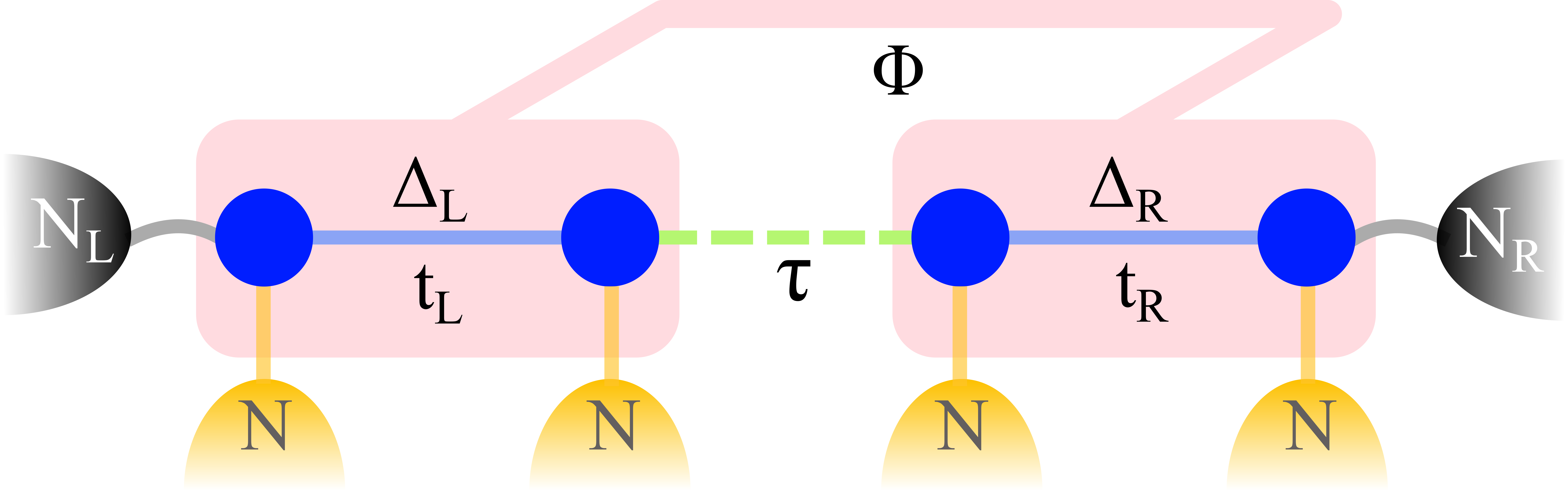}
\caption{A phase-biased Josephson junction formed by two minimal Kitaev chains (red boxes). The quantum dots (blue filled circles) in the left (L) and right (R) minimal Kitaev chain are coupled by electron tunneling $t_{\alpha}$ and $p$-wave pair potential $\Delta_{\alpha}$ characterizing crossed Andreev reflections, here $\alpha={\rm L,R}$. Also, the left and right minimal Kitaev chains are connected by regular electron tunneling $\tau$ and the superconducting phase difference between their pair potentials is controlled   by an external   magnetic flux $\Phi$ applied through the SQUID loop (red arm). All the quantum dots are coupled to normal reservoirs (N, yellow) which are sources   of non-Hermitian effects, while   the first (second) quantum dot  in the left (right) minimal  Kitaev chain is also weakly coupled to an extra normal reservoir $N_{\rm L(R)}$ for assessing transport.}
\label{Fig1} 
\end{figure}

In this work, we consider a Josephson junction formed by two coupled minimal Kitaev chains [Fig.\,\ref{Fig1}] and investigate the emergence of   non-Hermitian topological phases characterized by stable EPs. To be more precise, we focus on non-Hermiticity arising due to coupling to normal leads, as it realistically occurs when assessing quantum transport \cite{datta1997electronic}. We find that, depending on the spatial profile of non-Hermiticity, provided it is not homogeneous, the non-Hermitian system exhibits a complex Andreev spectrum with second-order EPs occurring either at zero or finite real energies and fully controlled by the superconducting phase difference between the minimal Kitaev chains. These Andreev EPs are protected by non-Hermitian topology and connect stable energy lines that characterize   emergent non-Hermitian topological phases. We then show that, when varying other parameters of the system, such as onsite energies, coupling to the leads, or the ratio between ECT and CARs, the Andreev EPs can appear along lines, giving rise to stable Andreev exceptional lines. Lastly, we show that local and nonlocal conductances can be used to identify the formation of Andreev EPs and their connecting zero real energy lines, with the local conductance being more beneficial at the sweet spot.

\section{Non-Hermitian Josephson junctions}
\label{SectionII}
We consider a non-Hermitian Josephson junction formed by two minimal Kitaev chains, as sketched in Fig.\,\ref{Fig1}. The  low-energy spectral properties of this non-Hermitian system can be  determined by the retarded Green's function
\begin{equation}
\mathcal{G}_{\rm JJ}^{r}(\omega,\phi)=[\omega-H_{\rm eff}(\omega,\phi)]^{-1}
\end{equation}
where $H_{\rm eff}(\omega,\phi)$ is the effective Hamiltonian modeling the open Josephson junction and given by
 \begin{equation}
\label{Eqeff}
H_{\rm eff}(\omega,\phi)=H_{\rm JJ}(\phi)+\Sigma^{r}(\omega)\,.
\end{equation}
Here,   $H_{\rm JJ}(\phi)$ describes the Hermitian Josephson junction with a superconducting phase difference $\phi$, and $\Sigma^{r}(\omega)$ is the retarded self-energy that incorporates the non-Hermitian effects due to coupling to normal reservoirs, see Fig.\,\ref{Fig1}. The Hermitian Josephson junction   is modeled by
 \begin{equation}
  H_{\rm JJ}(\phi_{R}-\phi_{L})=H_{\rm L}(\phi_{L})+H_{\rm R}(\phi_{R})+H_{\rm T}
\end{equation}
where $H_{\rm \alpha}$, with $\alpha=L,R$, is the Hamiltonian of each minimal Kitaev chain,  
and $H_{\rm T}$ is the tunneling coupling between the chains given by
\begin{equation}
\begin{split}
H_{\alpha}(\phi_{\alpha})&=\varepsilon_{\alpha_{i}}c^{\dagger}_{\alpha_{1}}c_{\alpha_{1}}+\varepsilon_{\alpha}c^{\dagger}_{\alpha_{2}}c_{\alpha_{2}}\\
&+t_{\alpha} c^{\dagger}_{\alpha_{1}}c_{\alpha_{2}}+\Delta_{\alpha} c_{\alpha_{1}}c_{\alpha_{2}}+{\rm h.\,c.}\,,\\
H_{\rm T}&=\tau c^{\dagger}_{\rm L_{2}}c_{\rm R_{1}}+{\rm h.\,c.}
\end{split}
\end{equation}
Here, $c^{\dagger}_{\alpha_{i}} (c_{\alpha_{i}})$ creates (annihilates) an electronic state in the chain $\alpha=L,R$ at quantum dot (QD) $i=1,2$, $\varepsilon_{\alpha_{i}}$ is the onsite energy of each QD,  and $t_{\alpha}$ is the hopping amplitude between QDs in each chain $\alpha$, while $\Delta_{\alpha}=\Delta {e}^{i\phi_{\alpha}}$ is the $p$-wave pair potential  with the superconducting phase $\phi_{\alpha}$ of   chain $\alpha$. We emphasize that $t_{\alpha}$ and $\Delta_{\alpha}$ characterize ECT and CAR processes between QDs, respectively. The control of ECT and CAR has already been demonstrated   in  minimal Kitaev chains fabricated by connecting two QDs via a superconductor-semiconductor hybrid system \cite{dvir2023realization,PhysRevX.13.031031,bordin2023crossed}.

Regarding the self-energy $\Sigma^{r}(\omega)$ in Eq.\,(\ref{Eqeff}) due to the coupling to normal reservoirs, our setup contains two contributions: one self-energy term coming from the normal reservoirs (N) located below the junction (yellow) in Fig.\,\ref{Fig1}, and another contribution due to the coupling to the left and right normal reservoirs (N$_{\rm L,R}$) located on the left and right sides of the junction (black). Under general circumstances, the self-energies in both cases have real and imaginary parts \cite{datta1997electronic,PhysRevB.105.094502}: while the real (Re) parts renormalize the   Hermitian Hamiltonian, the imaginary (Im) parts render the system non-Hermitian and introduce intriguing non-Hermitian topological features \cite{JorgeEPs,PhysRevB.105.094502,cayao2023nonhermitian,cayao2024non,cayaoSatoNH4MZMs}. Moreover, having imaginary self-energies is also useful for  assessing transport in the considered Josephson junction \cite{datta1997electronic}. In this regard, by coupling each QD of the minimal Kitaev chains to a normal reservoir [Fig.\,\ref{Fig1}], we can explore the effects of non-Hermiticity, whereas by coupling to the left and right normal reservoirs, we can assess transport signatures via, e.g., conductance \cite{datta1997electronic}.  Since we aim to inspect the impact of non-Hermiticity on the Josephson junction, in what follows we consider imaginary self-energies and, within the wide-band approximation \cite{datta1997electronic,PhysRevB.105.094502}, we also neglect their frequency dependence.  We therefore characterize the non-Hermitian effect of the normal reservoirs by a zero-frequency retarded self-energy,  
 \begin{equation}
 \label{SigmaTotal}
\begin{split}
\Sigma^{r}(\omega=0)&=-i \sum_{j,\alpha}  \Gamma_{\alpha_{j}} c^{\dagger}_{\alpha_{j}}c_{\alpha_{j}}\\
&-i\Gamma_{\rm L}c^{\dagger}_{\rm L_{1}}c_{\rm L_1}-i\Gamma_{\rm R}c^{\dagger}_{\rm R_{2}}c_{\rm R_2}\,,
\end{split}
\end{equation}
where $\Gamma_{\alpha_{j}}=\pi|t'_{\alpha_{j}}|^{2}\rho_{\alpha_{j}}$, with $\rho_{\alpha_{j}}$ being the surface density of states of the lead $\alpha$ at QD $j$, and $t'_{\alpha_{j}}$ the hopping amplitude between the leads and QDs. Since transport experiments in minimal Kitaev chains involve coupling to normal leads \cite{dvir2023realization,PhysRevX.13.031031,bordin2023crossed}, the control over the couplings  $\Gamma_{\alpha_{j}}$ is within experimental reach. We stress that the couplings $\Gamma_{\rm L,R}$ are considered to be weak since they will be used to assess conductance, whereas the values of $\Gamma_{\rm \alpha_{j}}$ are considered to be stronger since they are required for inducing non-Hermitian effects.

\subsection{Particle-hole symmetry and non-Hermitian topology in Josephson junctions}
\label{SectionIIa}
Before proceeding, following our first study on non-Hermitian Josephson junctions \cite{cayao2023nonhermitian}, we briefly discuss the particle-hole symmetry of the effective Josephson junction Hamiltonian $H_{\rm eff}$ [Eq.\,(\ref{Eqeff})] and its non-Hermitian topology. 

We start by inspecting the particle-hole symmetry (PHS) of regular Hamiltonians in Nambu space. These Hamiltonians are quadratic and in Nambu space can be written as
\begin{equation}
\label{quadraticH}
\hat{H}=\sum_{\alpha i,\beta j}\psi^\dagger_{\alpha i}(H)_{\alpha i,\beta j}\psi_{\beta j}\,,  
\end{equation}
where $\psi_{\alpha i}$ is the spinor in Nambu space, $\alpha$ denotes   the JJ ($\alpha=1$) and the reservoir ($\alpha=2$), while $i$ represents additional degrees of freedom such as the coordinate and spin. To examine the role of particle-hole symmetry (PHS), it is important to recall that PHS is a proper symmetry of superconductors and is defined as,
\begin{equation}
\label{Seq:PHS}
\begin{split}
\hat{C}\psi^\dagger_{\alpha i} \hat{C}^{-1}&=\sum_{j}\psi_{\alpha j}(U^\alpha_C)^*_{ji}\,,\\
\hat{C}\psi_{\alpha i} \hat{C}^{-1}&=\sum_{j}\psi^\dagger_{\alpha j}(U^\alpha_C)_{ji}\,,
\end{split}
\end{equation}
where   $U^\alpha_C$ is a unitary matrix and $\hat{C}$ is a unitary operator. We stress that PHS is a local symmetry and, therefore, it does not mix different $\alpha$. Then, applying the operations from Eqs.\,(\ref{Seq:PHS}) to the Hamiltonian given in Eq.\,(\ref{quadraticH}), one reproduces the standard form of PHS
\begin{equation}
\label{Seq:PHS2}
[\hat{H},\hat{C}]=0   \Leftrightarrow 
   (U_C^\alpha)_{ip}(H^T)_{\alpha p, \beta q} (U_C^{\beta\dagger})_{q j}=-(H)_{\alpha i, \beta j}\,.
\end{equation}

To assess PHS in our effective Hamiltonian, we inspect  the associated retarded Green's function
of the non-Hermitian JJ
\begin{align}
(\mathcal{G}^{\rm r}_{\rm JJ})_{ij}=-i\theta(t)\langle \psi_{1i}(t)\psi^\dagger_{1j}+\psi^\dagger_{1j}\psi_{1i}(t)\rangle,
\end{align} 
where $\psi_{1i}$ is the Nambu spinor in the JJ. We then use the fact that PHS dictates  that
 $\hat{C}^{-1}\hat{H}\hat{C}=\hat{H}$, leading to
\begin{equation}
\langle \psi_{1i}(t)\psi^\dagger_{1j}\rangle=\sum_{kl}(U^1_C)^\dagger_{jl}
\langle \psi_{1l}^\dagger \psi_{1k}(t)\rangle^* (U^1_{C})_{ki}\,,
\label{Seq:aa}
\end{equation}
where we have used the Hermiticity of the total Hamiltonian $\hat{H}$ including both JJ and the reservoir. Therefore, for the Green's function, we obtain 
\begin{equation}
\label{Geff}
(\mathcal{G}^{\rm r}_{\rm JJ})_{ij}=-(U_C^1)^\dagger_{jl}(\mathcal{G}^{\rm r}_{\rm JJ})^{*}_{kl}(U_C^1)_{ki}\,.
\end{equation}
Thus, using Eq.\,(\ref{Geff}) one can  directly obtain the particle-hole symmetry in the effective Hamiltonian of the JJ, which reads
\begin{equation}
\label{Seq:PHSdagger}
(U_C^1)^T H_{\rm eff}^* (-\omega, \phi)(U_C^1)^*=-H_{\rm eff}(\omega,\phi)\,,
\end{equation}
where   $\phi$ is the phase difference across the JJ. Importantly,  PHS in the effective Hamiltonian involves complex conjugation, in contrast to the original PHS in Eq.(\ref{Seq:PHS2}). This complex-conjugation version of PHS is  called PHS$^{\dagger}$ and directly defines distinct non-Hermitian topological phases \cite{PhysRevX.9.041015}. For instance, since $U^\alpha_C$ is unitary,  when $U^\alpha_C(U^\alpha_C)^{*}=1$, PHS$^{\dagger}$ defines class $D^{\dagger}$ in the symmetry classification which exhibits a $0-$dimensional $\mathbb{Z}_{2}$  point-gap topological number \cite{PhysRevLett.123.066405}. Ref.\,\cite{cayao2023nonhermitian}  showed that  EPs at zero real energy possess a nontrivial $\mathbb{Z}_{2}$ point-gap topological number in Josephson junctions based on conventional superconductors. These EPs are robust   since PHS$^{\dagger}$ in class $D^{\dagger}$ remains intact under any perturbation in superconducting systems, provided the superconducting gap stays open or EPs are not pair-annihilated. 
It is therefore possible to engineer stable non-Hermitian   topological phases by exploiting non-Hermitian topology and Hermitian Josephson junctions.

\section{Andreev exceptional points}
We now inspect the formation of exceptional points by means of the superconducting phase difference, which, in the spirit of previous works  \cite{cayao2023nonhermitian,li2023anomalous,cayao2024non,cayaoSatoNH4MZMs}, we refer to    as Andreev exceptional points (Andreev EPs). For this purpose, we obtain the energy spectrum of the effective non-Hermitian Hamiltonian $H_{\rm eff}(\omega,\phi)$ derived from the retarded Green's function   given by Eq.\,(\ref{Eqeff}).  Thus, the poles of the Green's function define the energies, which are solutions to ${\rm det}\{[\mathcal{G}^{r}(\omega,\phi)]^{-1}\}={\rm det}(\omega-H_{\rm eff}(0,\phi))\equiv D(\omega,\phi)=0$ \cite{mahan2013many,zagoskin}. Under general conditions, $D(\omega,\phi)$ is given by
\begin{equation}
\label{De}
D(\omega,\phi)=\sum_{i=0}^{8}P_{i}(\phi)\,\omega^{i}
\end{equation}
where $P_{8}(\phi)=1$ and the rest $P_{i}(\phi)$ are lengthy expressions that depend on all the system parameters . Owing to particle-hole symmetry, the solutions to $D(\omega,\phi)=0$ come in pairs, but their general expressions do not have a simple form. Furthermore, because of the non-Hermiticity introduced by the couplings to the reservoirs $\Gamma_{\alpha_j}$ in $P_{i}(\phi)$, the spectrum is expected to be complex; here, we set $\Gamma_{\rm L,R}=0$  in Eq.\,(\ref{SigmaTotal}) as they  are only relevant for assessing conductance signatures.  In order to understand the formation of Andreev exceptional points, in the following we discuss the energy spectrum in both the Hermitian and non-Hermitian regimes based on Eq.\,(\ref{De}).

 \begin{figure}[!t]
\centering
\includegraphics[width=0.49\textwidth]{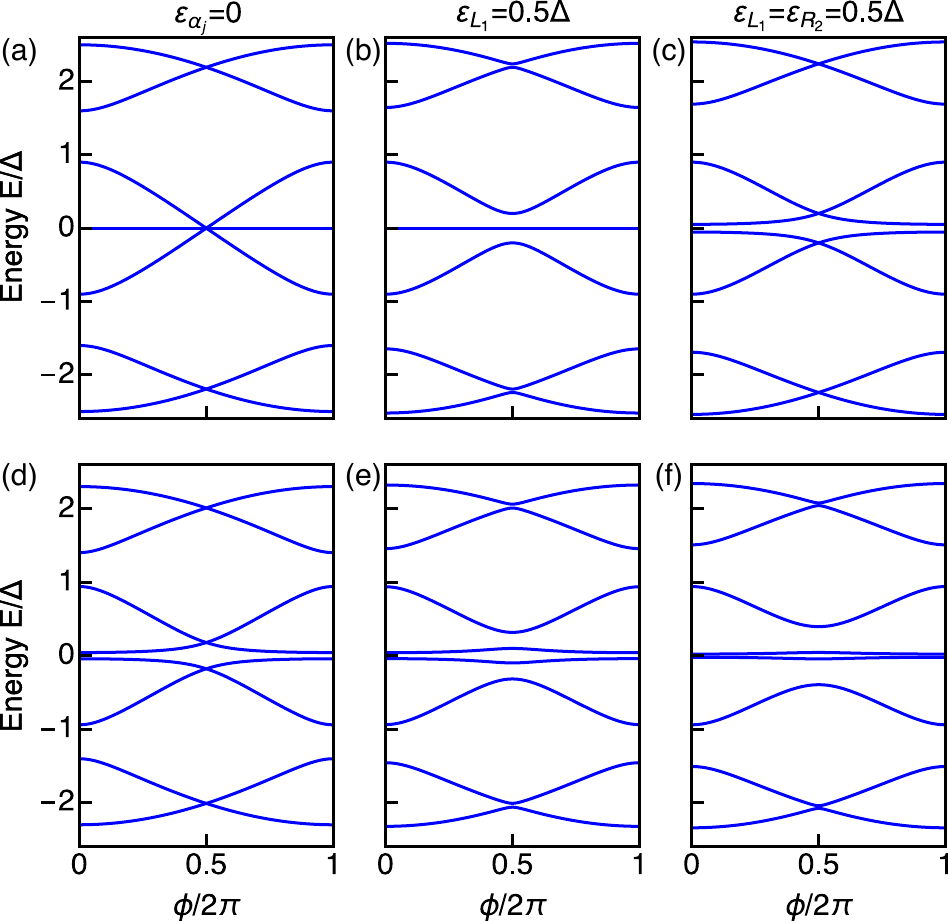}
\caption{Energy spectrum as a function of the superconducting phase difference $\phi$ in the Hermitian regime, where (a-c) and (d-f) correspond, respectively, to $t_{\alpha}=\Delta_{\alpha}$ and $t_{\alpha}=0.8\Delta_{\alpha}$ for distinct values of the onsite energies $\varepsilon_{\alpha_j}$. In the left column (a,d), all onsite energies are equal to zero $\varepsilon_{\alpha_j}=0$; in the middle column (b,e), only the onsite energy of the leftmost QD is finite ($\varepsilon_{\rm L_1}\neq0$); in the right column (c,f), only the energies of the outer QDs are finite ($\varepsilon_{\rm L_1}=\varepsilon_{\rm R_2}\neq0$). Parameters: $\tau=0.9\Delta$, $\Delta_{\alpha}\equiv\Delta$}
\label{Fig2} 
\end{figure}

\subsection{Andreev spectrum in the Hermitian regime}
In the Hermitian regime, the system is closed and characterized by $\Gamma_{\alpha_j}=0$. The polynomial equation given by Eq.\,(\ref{De}) acquires a simpler form and, at $\Delta_{\alpha}=t_{\alpha}$ and $\varepsilon_{\alpha}=\varepsilon$, it becomes
\begin{equation}
\label{DP}
D(\omega,\phi)=\omega^{8}+P_{6}\omega^{6}+P_{4}\omega^{4}+P_{2}(\phi)\omega^{2}+P_{0}\,,
\end{equation}
where $P_{0}=\varepsilon^{4}(\varepsilon^{4}-2\varepsilon^{2}\tau^{2}+\tau^{4})$, $P_{2}=2\varepsilon^{4}(-2\varepsilon^{2}-4t^{2}\tau^{2})-2\varepsilon^{2}\tau^{2}(4t^{2}+\tau^{2})-16t^{4}\tau^{2}{\rm cos}^{2}(\phi/2)$, $P_{4}=6\varepsilon^{4}+(4t^{2}+\tau^{2})^{2}+2\varepsilon^{2}(8t^{2}+\tau^{2})$, $P_{6}=-4\varepsilon^{2}-2(4t^{2}+\tau^{2})$. From these expressions, we see that, at $\varepsilon=0$, $P_{0}=0$, $P_{2}(\phi)=-16t^{4}\tau^{2}{\rm cos}^{2}(\phi/2)$, $P_{4}=(4t^{2}+\tau^{2})^{2}$, and $P_{6}=-2(4t^{2}+\tau^{2})$, meaning the energy spectrum is   obtained by solving  

\begin{equation}
\label{ABSHermitianSS}
\omega^{2}\{\omega^{2}[\omega^{2}-(4t^{2}+\tau^{2})]^{2}-16t^{4}\tau^{2}{\rm cos}^{2}(\phi/2)\}=0\,.
\end{equation}
Thus, in this sweet-spot regime ($\varepsilon=0$ and $\Delta_{\alpha}=t_{\alpha}$), Eq.\,(\ref{ABSHermitianSS}) reveals that    two solutions appear at $\omega_{\pm}^{(0)}=0$ irrespective of the $\phi$ and other parameters; these zero-energy states represent dispersionless Majorana states located at the outer QDs of each Kitaev chain and shown in Fig.\,\ref{Fig2}(a). 
Eq.\,(\ref{ABSHermitianSS}) also shows that the Hermitian Josephson junction hosts energy levels that depend on $\phi$, see the terms inside the curly brackets of that equation. By looking at the low-frequency solutions, we obtain $\omega_{\pm}^{(1)}(\phi)=f(t,\tau){\rm cos}(\phi/2)$, with $f(t,\tau)=2\sqrt{2}\tau t^{2}/(4t^{2}+\tau^{2})$: these ABSs   disperse with $\phi$ and acquire zero energy at $\phi=\pi$ [Fig.\,\ref{Fig2}(a)], becoming Majorana states located at the inner QDs.  The zero-energy crossing at $\phi=\pi$ is protected by a local charge-conjugation symmetry, inherent to minimal Kitaev chains \cite{cayaosatosymmetryMKC,hf7s-f7tj}.  Besides the low-energy sector, the junction in this sweet-spot regime also hosts dispersing  levels at higher energies, which develop a crossing at $\phi=\pi$.  

When $\varepsilon_{\alpha_j}\neq0$ but still $\Delta_{\alpha}=t_{\alpha}$, several changes occur in the spectrum that are not   captured by the solutions of Eq.\,(\ref{ABSHermitianSS}); instead, the energy spectrum  requires  solving Eq.\,(\ref{DP}). The resulting solutions are presented in Fig.\,\ref{Fig2}(b) as a function of $\phi$ for a nonzero onsite energy of the leftmost outer QD $\varepsilon_{\rm L_1}$ at $\Delta_{\alpha}=t_{\alpha}$.  In this case, the ABSs that tend to cross zero energy at $\phi=\pi$ develop a finite energy, but the outer MZMs still remain at zero energy. The finite energies of the ABS originate from  the fact that their wavefunctions hybridize because they are not entirely localized in the inner QDs, while  the levels remaining at zero energy continue to be located at the outer QDs; this behavior persists even if the energy of another QD in the same chain is nonzero. However, when the energies of the outer QDs are nonzero in Fig.\,\ref{Fig2}(c), the inner and outer Majorana states split at $\phi=\pi$ due to a strong hybridization of their wavefunctions.  In this case, the outer MZMs do not remain at zero anymore. It is worth noting that the low-energy sector discussed here resembles the formation of four Majorana states in Josephson junctions with finite length topological superconductors \cite{PhysRevLett.108.257001,PhysRevB.86.140504,PhysRevB.91.024514,PhysRevB.89.014509,PhysRevB.96.205425,PhysRevB.96.165415,cayao2018andreev,PhysRevB.94.085409,PhysRevLett.123.117001,PhysRevB.104.L020501}, where their energy splitting is controlled by the ratio between the length of the superconductor and the Majorana localization length. Away from the sweet spot, when $\Delta_{\alpha}\neq t_{\alpha}$, the lowest  levels always exhibit  finite energy as a result of the hybridization of their wavefunctions, showing that this regime does not host Majorana states.

\subsection{Non-Hermitian Andreev spectrum at $\Delta_{\alpha}=t_{\alpha}$}
We are now in a position to discuss the effect of non-Hermiticity on  the spectrum of our non-Hermitian Josephson junction described by Eq.\,(\ref{Eqeff}). In this case, non-Hermiticity renders the spectrum complex, and it is obtained  by solving $D(\omega,\phi)=0$ for $\omega$, with $D(\omega,\phi)$ given by Eq.\,(\ref{De}). Under general conditions, when all the couplings are equal ($\Gamma_{\alpha_j}\equiv\Gamma$), the complex spectrum consists of a real (Re) part and an imaginary (Im) part that is constant and equal to $-i\Gamma$. When there is an asymmetry in the couplings $\Gamma_{\alpha_j}$, the   complex spectrum  becomes more interesting, with properties that are entirely due to non-Hermiticity.  We start by looking at the regime with $\Delta_{\alpha}=t_{\alpha}$ and equal non-Hermiticity in each QD of  the left chain, $\Gamma_{\rm L_1}=\Gamma_{\rm L_2}\neq0$, while we set   the couplings of the QDs in the right chain to zero. In this regime, fully solving $D(\omega,\phi)=0$ is still challenging, but it is  possible to first analytically obtain  the lowest energies closest to zero  by taking $D(\omega,\phi)$ up to second order in $\omega$, which, at $\varepsilon_{\alpha_j}=0$, read
\begin{equation}
\label{solABS0}
\begin{split}
\omega^{(0)}_{+}(\phi)&=0\\
\omega_{-}^{(0)}(\phi)&=-i\frac{4\Gamma \tau^{2}\Delta^{2}[\Gamma^{2}+4\Delta^{2}{\rm cos}^{2}(\phi/2)]}{\Gamma^{2}P+4\tau^{2}\Delta^{2}[\Gamma^{2}+4\Delta^{2}{\rm cos}^{2}(\phi/2)]}\,,
\end{split}
\end{equation}
where $P=\left[4\Gamma_{\rm L_{2}}^{2}\Delta^{2}+(\tau^{2}+4\Delta^{2})\right]$,  $\Gamma=\Gamma_{\rm L_{1,2}}$, and $\Delta_{\alpha}=t_{\alpha}\equiv\Delta$. These expressions reveal that the lowest energies in this sweet-spot regime ($\Delta_{\alpha}=t_{\alpha}$ and $\varepsilon_{\alpha_j}=0$) exhibit zero Re energy even under the effect of non-Hermiticity characterized by $\Gamma_{\rm L_1}=\Gamma_{\rm L_2}\neq0$; we recall that in the Hermitian regime, these states are located at the outer QDs. Interestingly, these lowest energies develop distinct Im parts, with $\omega^{(0)}_{+}(\phi)$ having a zero Im part, while $\omega^{(0)}_{-}(\phi)$ is completely imaginary and dependent on $\phi$. This means that $\omega^{(0)}_{+}(\phi)$ remains localized at one outer QD, while $\omega^{(0)}_{-}(\phi)$ becomes delocalized, leaking into the respective leads as it acquires a finite lifetime ($\sim 1/{\rm Im}[\omega^{(0)}_{-}(\phi)]$), which is the time a quasiparticle spends in the chain before escaping into the normal leads.

 \begin{figure}[!t]
\centering
\includegraphics[width=0.49\textwidth]{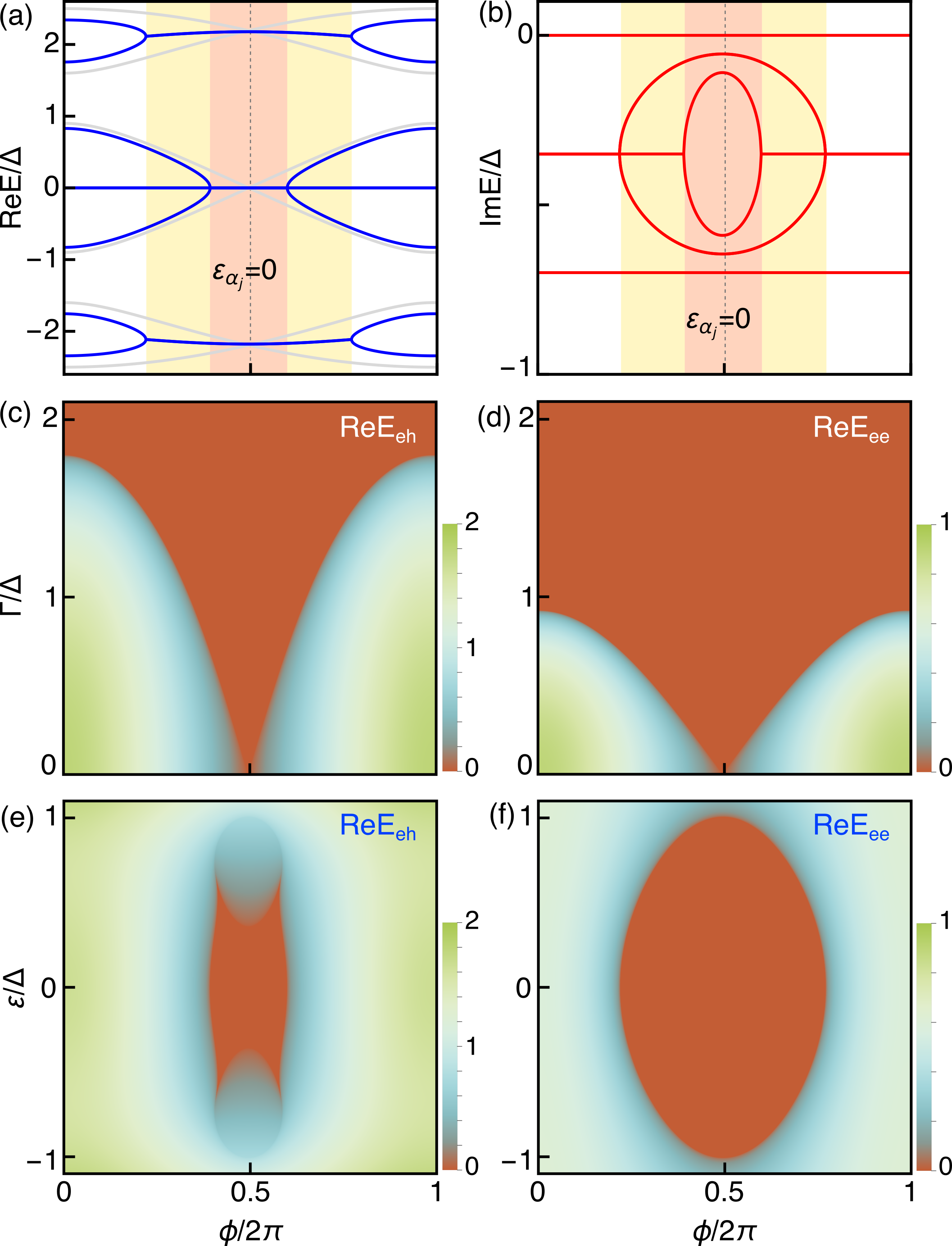}
\caption{(a,b) Energy spectrum versus $\phi$ in the non-Hermitian regime ($\Gamma_{\rm L_j}\neq0$, $\Gamma_{\rm R_j}=0$) at $\Delta_{\alpha}=t_{\alpha}$ and $\varepsilon_{\alpha_j}=0$, with (a) and (b) showing the real (Re) and imaginary (Im) parts, respectively.  The ends of the orange and yellow shaded regions mark the position of Andreev EPs emerging at zero and finite Re energies, while the gray curves in (a) correspond to the energies in the Hermitian regime. (c) Real part of the difference between the first positive and first negative excited   levels closest to zero Re energy versus $\phi$ and $\Gamma\equiv  \Gamma_{\rm L_j}$. (d) Same as (a) but for the highest two positive levels. (e,f) Same quantity as in (c,d) as a function of $\phi$ and onsite energy $\varepsilon\equiv \varepsilon_{\alpha_j}$.  Parameters: $\tau=0.9$, $\Delta_{\alpha}\equiv\Delta$, $\varepsilon_{\alpha_j}=0$ in (c,d), $\Gamma\equiv \Gamma_{\rm L_j}=0.7\Delta$ in (a,b,e,f), $\Gamma_{\rm R_j}=0$.}
\label{Fig3} 
\end{figure}

The above discussion can be  visualized by numerically solving $D(\omega,\phi)=0$ for the considered regime; the Re and  and Im energies are  plotted  in Fig.\,\ref{Fig3}(a,b) as a function of $\phi$. Here, we confirm the discussion of the previous paragraph at the sweet spot: two levels at zero Re energy appear  corresponding to the outer MZMs;  one of these states has a zero Im part and the other has a finite Im part,  whose phase dependence  in the full solution is negligible and whose value is determined by  $-i\Gamma_{\rm L_2}$ (as predicted by the second expression of Eqs.\,(\ref{solABS0})).  Moreover, the first excited positive and negative energy levels, which in the Hermitian regime develop a zero-energy crossing at $\phi=\pi$  and characterize the inner MZMs, merge into  a flattened zero Re energy line around $\phi=\pi$, while their Im parts  split in this regime, see the red shaded region in Fig.\,\ref{Fig3}(a).  This implies that the Hermitian point  with inner MZMs expands into a larger region under non-Hermiticity, which can be seen as beneficial for the detection of MZMs. We verified that the ends of the zero Re energy line correspond to the formation of exceptional points (EPs), where the eigenvalues and  their corresponding eigenfunctions coalesce \cite{el2018non,ozdemir2019parity,doi:10.1080/00018732.2021.1876991,heiss2004exceptional,Heiss_2012}. Unlike Hermitian degeneracies, the energy behavior around such EPs is not linear but instead follows a $(\phi-\phi_{\rm EP})^{1/2}$ dispersion,  where  the  $2$ in the exponent $1/2$ indicates the second order nature of the obtained EP \cite{doi:10.1080/00018732.2021.1876991}. As a result, the responsivity $1/\sqrt{\phi-\phi_{\rm EP}}$ exhibits a divergent behavior, which is distinct from the finite constant responsivity expected  in the Hermitian regime. Furthermore, the choice of non-Hermiticity ($\Gamma_{\rm L_2}=\Gamma_{\rm L_1}$) in this sweet-spot regime leads to EPs at higher  Re energies, which originate from  the coalescence of the two positive (negative) energy levels around $\phi=\pi$:  These higher Re energy EPs connect a finite Re energy line that is weakly dependent on $\phi$, while their Im parts split in a circular fashion, see the yellow shaded area in Fig.\,\ref{Fig3}(a,b). This finite Re energy line stands, of course, in contrast to the constant zero Re energy line connecting EPs at zero Re energy. 

While both zero and finite Re energy EPs emerge as a result of $\phi$, and hence due to the Josephson effect, their emergence can be controlled by the other system parameters. To show these features, in Fig.\,\ref{Fig3}(c,d) we present the differences between the Re parts of the energy levels giving rise to EPs at zero and finite Re energies (denoted as ${\rm Re}E_{\rm eh}$ and ${\rm Re}E_{\rm ee}$) as functions of $\phi$ and $\Gamma_{\rm L_{1,2}}=\Gamma$, both at the sweet spot. The first feature is that  both ${\rm Re}E_{\rm eh}$ and  ${\rm Re}E_{\rm ee}$ reveal EPs, which correspond to the borders of the orange region and thus signal a line of EPs, or an exceptional line, driven by $\Gamma$ and $\phi$. Note that, despite their similar behavior of ${\rm Re}E_{\rm eh}$ and ${\rm Re}E_{\rm ee}$, the finite Re energy EPs signaled by  ${\rm Re}E_{\rm ee}$ disappear at lower values of $\Gamma$ than those at zero Re energy revealed by ${\rm Re}E_{\rm eh}$. While these  EPs  are obtained at $\varepsilon_{\alpha_j}=0$, they remain robust at finite and equal onsite energies, with the zero Re energy regions in each case developing  distinct profiles, as shown in  Fig.\,\ref{Fig3}(e,f). We have verified  that the zero Re energy EPs remain robust against distinct and nonzero onsite energies, whereas those at finite Re energies are more fragile although they still form when the onsite energies of either the inner or outer QDs are equal and finite.

 \begin{figure}[!t]
\centering
\includegraphics[width=0.49\textwidth]{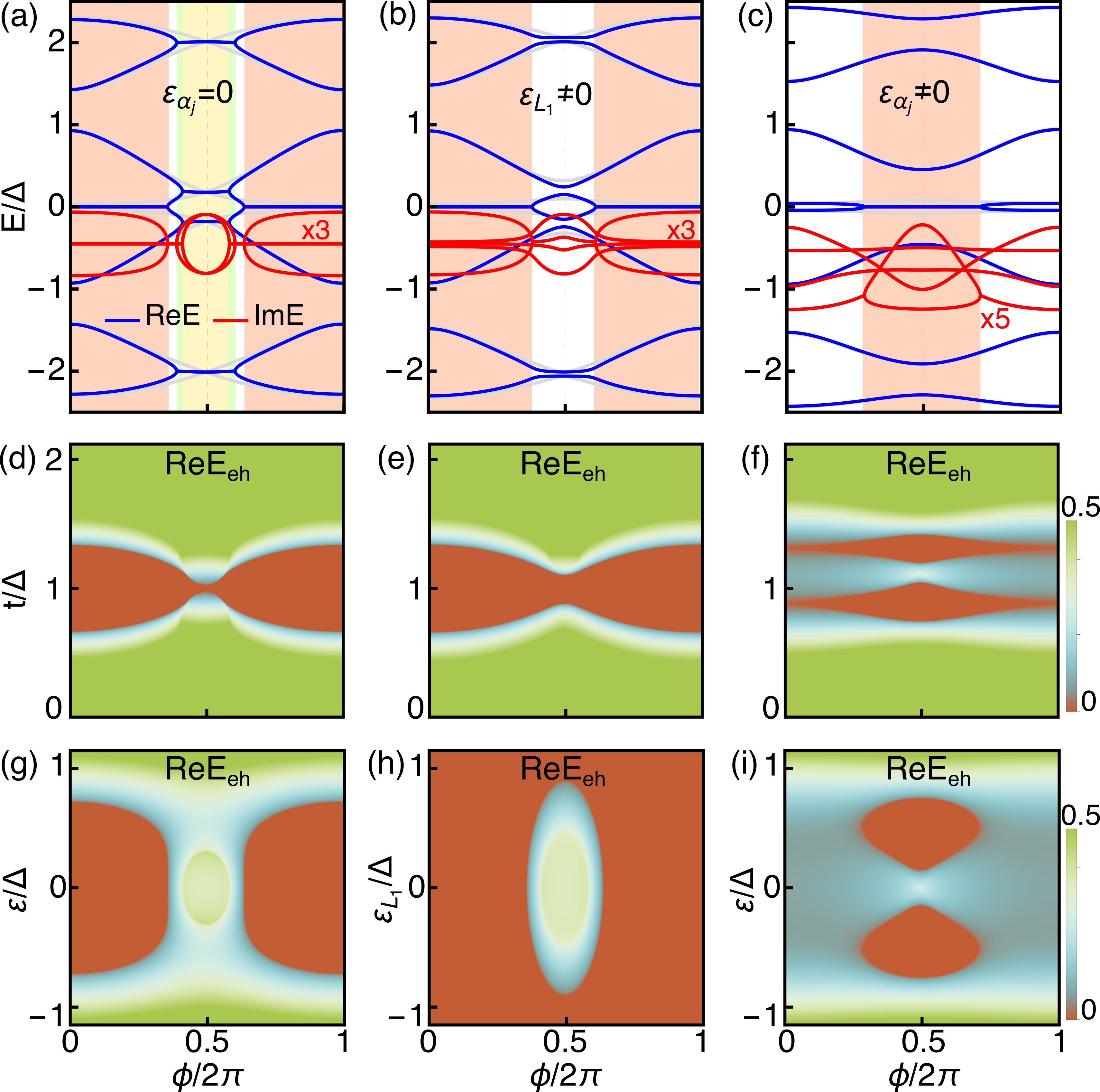}
\caption{(a,b) Real (blue) and imaginary (red) parts of the energy spectrum versus $\phi$ in the non-Hermitian regime for   $\Delta_{\alpha}\neq t_{\alpha}$ at distinct values of the onsite energies and couplings:  (a,b) for   $\Gamma_{\rm L_j}\neq0$, $\Gamma_{\rm R_j}=0$, while (c) for $\Gamma_{\rm L_1,R_2}\neq0$, $\Gamma_{\rm L_2,R_1}=0$. The ends of the red, yellow, and green shaded regions mark the positions of the Andreev EPs emerging due to the coalescence of lowest positive and lowest negative levels, lowest positive (negative) and first excited positive (negative) levels, and highest positive (negative) and highest positive (negative) levels, respectively. (d-f) Real part of the difference between the first positive and first negative excited levels closest to zero Re energy versus $\phi$ and $t/\Delta$ for the onsite energies and $\Gamma_{\alpha_j}$ of (a-c), respectively; here, the color scale is cut off at 0.5 for visualization. (g-i) The same quantity as in (d-f) but as a function of $\phi$ and $\varepsilon\equiv\varepsilon_{\alpha_j}$ (g), $\varepsilon_{\rm L_1}$ (h), and     $\varepsilon\equiv\varepsilon_{\alpha_j}$.   Parameters: $\tau=0.9$, $\Delta_{\alpha}\equiv\Delta$, $\varepsilon_{\rm L_1}=0.5\Delta$ in (b,e), $\varepsilon=0.5\Delta$ in (c,f),  $\Gamma_{\rm L_j}=0.3\Delta$ in (a,b,d,e,g,h), $\Gamma_{\rm L_1,R_{2}}=0.3\Delta$ in (c,f,i).}
\label{Fig4} 
\end{figure}

\subsection{Non-Hermitian Andreev spectrum at $\Delta_{\alpha}\neq t_{\alpha}$}
We now explore the formation of Andreev EPs away from the sweet spot,  $\Delta_{\alpha}\neq t_{\alpha}$; we set $\Delta_{\alpha}=\Delta$ and $t_{\alpha}=t$. As in the previous subsection, we numerically obtain the complex spectrum by solving $D(\omega,\phi)=0$ for $\omega$  using Eq.\,(\ref{De}), and it is presented in Figs.\,\ref{Fig4}(a-c) as a function of $\phi$ for distinct values of the onsite energies ($\varepsilon_{\alpha_j}$) and couplings to the leads ($\Gamma_{\alpha_j}$). In particular, we assess regimes where only the QDs of the left chain exhibit equal nonzero non-Hermiticity ($\Gamma_{\rm L_1}=\Gamma_{\rm L_2}\neq0$) for  $\varepsilon_{\alpha_j}\equiv \varepsilon\neq0$ [Fig.\,\ref{Fig4}(a)] and $\varepsilon_{\rm L_1}\neq0$ [Fig.\,\ref{Fig4}(b)]; Fig.\,\ref{Fig4}(c) corresponds to $\Gamma_{\rm L_1}=\Gamma_{\rm R_2}\neq0$ and $\varepsilon_{\alpha_{j}}\equiv \varepsilon$. 
In this case, the particular profile of the Hermitian spectrum favors the formation of EPs at distinct Re energies.
In fact, at zero onsite energies, this regime hosts EPs at zero Re energy due to the coalescence between the lowest positive and lowest negative  levels; at finite but low Re energy due to the coalescence between the first excited level and the lowest level closest to zero;  and also at high energies between highest positive (negative) levels. The EPs at zero Re energy evolve from the relatively low energies in the Hermitian regime and connect a zero Re energy line for  $-\pi+2n\pi<\phi<\pi+2n\pi$ with $n\in\mathbb{Z}$, which are marked by the ends of the red shaded region in  Fig.\,\ref{Fig4}(a). The low but finite Re energy EPs
stem from the touching of levels at $\phi=\pi$ and are connected by a roughly constant finite Re energy around $\phi=\pi$, see 
the yellow shaded region in Fig.\,\ref{Fig4}(a). The EPs at the highest energies are similar to those found at the sweet spot in Fig.\,\ref{Fig3}(a) but the connecting Re energy line is shorter, see green shaded region in Fig.\,\ref{Fig4}(a). In all cases, the corresponding Im parts of the levels that coalesce  are split along the zero (finite) Re energy lines between EPs, with the EPs at finite Re energies developing a circular profile. 

The EPs at zero Re energy discussed above are robust against finite   onsite energies, while the finite Re energy EPs are strongly sensitive, see Fig.\,\ref{Fig4}(b). Indeed, the finite Re energy EPs can be robust against a symmetric profile of  QD onsite energies (either the same values at the inner or  outer QDs) but they disappear when such values are large or when their configuration is asymmetric, as   is the case shown in Fig.\,\ref{Fig4}(b) for a nonzero  onsite energy at the leftmost QD. By changing the non-Hermitian profile, it is possible to strongly affect the dependence of the energy levels, especially of those undergoing EP transitions.  For instance, inducing  a symmetric non-Hermiticity at the outer QDs of the Josephson junction can lead to EPs between  the first  positive and  first  negative  levels, which appear around $\phi=\pi$ and connect a zero Re energy line, see Fig.\,\ref{Fig4}(c). The  corresponding Im parts exhibit a  splitting for $\phi$ within the zero Re energy line, as seen for the zero Re energy EPs at the sweet spot, but now they develop a triangular shape dependence on $\phi$ with one of the corners tending to zero at $\phi=\pi$. The rest of the Im parts also develop a phase dependence even though they do not form EPs.  As already evident, the proliferation and robustness of EPs at zero Re energy makes them to possess a particular relevance for realizing zero-energy states even away from the sweet-spot regime.
 
To further explore the robustness of the zero Re energy EPs, in  Figs.\,\ref{Fig4}(d-f) we present the Re part of the difference  between the first positive and first negative energies (${\rm Re}E_{\rm eh}$) as a function  of $\phi$ and $t/\Delta$, with each panel corresponding to the onsite energies used in  Figs.\,\ref{Fig4}(a-c). We recall that $t$ and $\Delta$ characterize ECT and CAR processes, respectively. A common characteristic of all these regimes is the emergence of a zero Re energy region, ${\rm Re}E_{\rm eh}=0$, which strongly depends on the ratio between $t/\Delta$ and $\phi$. The borders of these zero Re energy regions correspond to a line of EPs, hence called exceptional line,   enclosing a broader region of zero energy states than in the Hermitian regime at the sweet spot: at the sweet spot ($t/\Delta=t$ and $\varepsilon_{\alpha_j}=0$), there is only one single line with ${\rm Re}E_{\rm eh}=0$ in Figs.\,\ref{Fig4}(d), which, results in an enlarged zero Re region even away from the sweet spot when   finite non-Hermiticity is present, as shown in Figs.\,\ref{Fig4}(d-f).  Note that while varying the onsite energies barely induces changes in the zero Re energy region [Figs.\,\ref{Fig4}(d,e)], it exhibits a noticeable dependence on the profile of non-Hermiticity [Figs.\,\ref{Fig4}(f)].  Furthermore, varying the onsite energies in each regime from Figs.\,\ref{Fig4}(a-c) introduces pronounced changes in the zero Re energy regions [Figs.\,\ref{Fig4}(g-i)], which  develop different shapes and   reveal the necessary onsite energies for achieving EPs. For instance, an equal amount of non-Hermiticity in each QD of the left chain immediately induces EPs at zero onsite energies which remain even when such energies take the same finite values. In contrast, when an equal non-Hermiticity is present in the outer QDs of the Josephson junction, the onsite energies need to be tuned to finite values. Thus, depending on the regime, EPs and zero energy states can be engineered at will by means of non-Hermiticity in minimal Kitaev chain Josephson junctions.

 Before closing this part, we would like to stress that the zero and finite Re energy EPs, and their connecting lines, discussed in this and previous subsections, are topologically protected  by distinct non-Hermitian symmetries. As discussed in SubSection \ref{SectionIIa}, the zero Re energy EPs are robust and protected by    ${\rm PHS}^{\dagger}$ symmetry in class $D^{\dagger}$  by a nontrivial $0$-dimensional  $\mathbb{Z}_{2}$ point-gap topological number \cite{PhysRevLett.123.066405}. Regarding the EPs at finite Re energies, they are protected by an accidental  parity-time (PT) symmetry, which enables a $0$-dimensional $\mathbb{Z}_{2}$ point-gap topological number \cite{PhysRevLett.123.066405} that ensures their stability  in the one-dimensional parameter space  of $\phi$.  We note that these two types of second order EPs are similar to those reported   in Ref.\,\cite{cayao2023nonhermitian}, which  utilized  Josephson junctions formed by conventional superconductors and non-Hermiticity originating from  ferromagnet leads. In our present study, we show that non-Hermiticity arising from  normal leads in Josephson junctions based on minimal Kitaev chains can able to induce stable phases protected by non-Hermitian topology.

\begin{figure}[!t]
\centering
\includegraphics[width=0.49\textwidth]{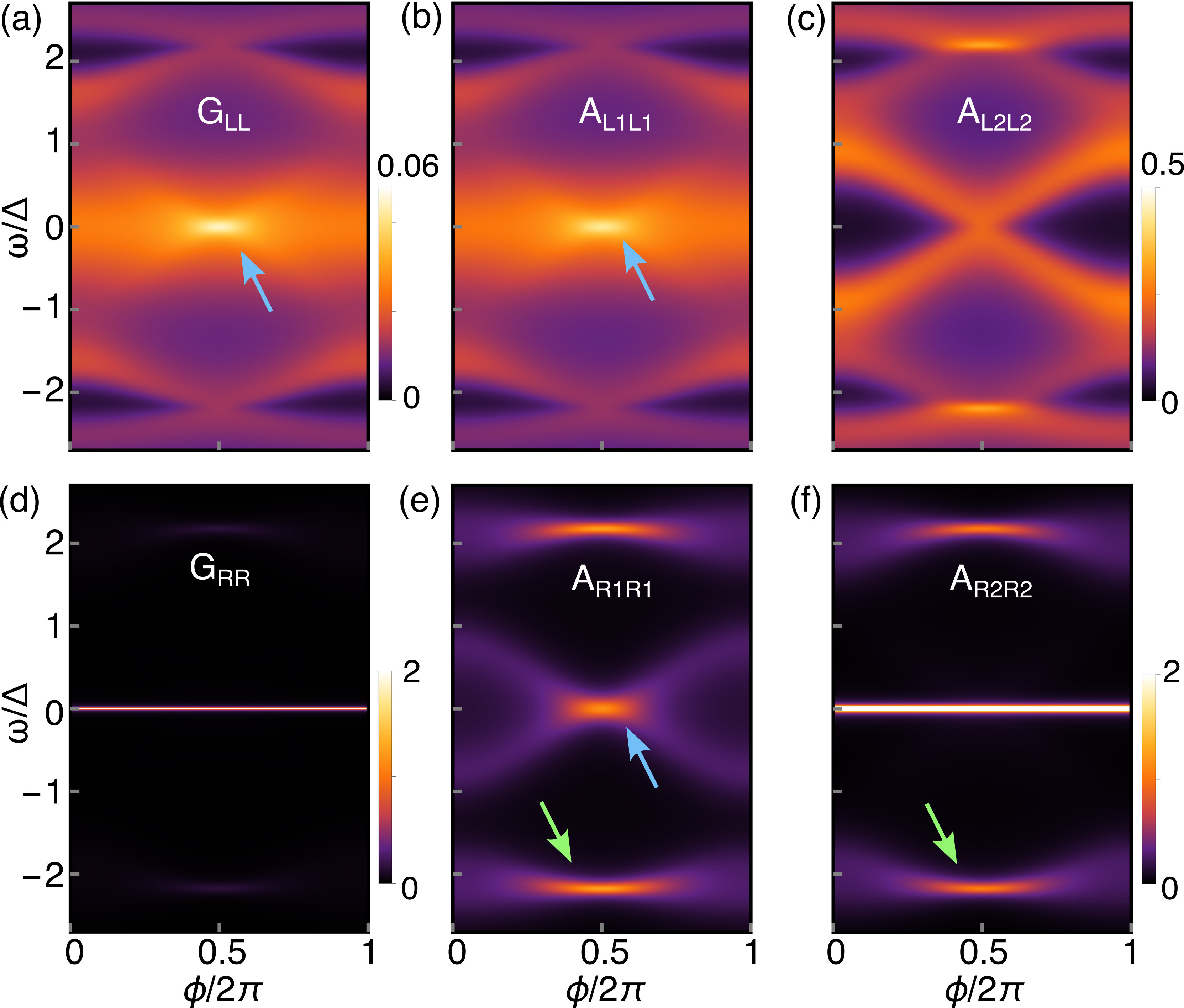}
\caption{Local conductance $G_{\alpha\alpha}$ (a,d) and spectral function $A_{\alpha_{j}\alpha_{j}}$ (b,c,e,f) as a function of frequency $\omega$ and $\phi$ at $\Delta_{\alpha}=t_{\alpha}$ and $\Gamma_{\rm L_j}\neq0$, $\Gamma_{\rm R_j}=0$. The conductance is given in units of $e^{2}/h$. In (e,f),  the color scale is cut off at 2 for visualization. The light blue (green) arrows indicate the appearance of the zero (finite) Re energy line connecting   EPs found at the sweet spot, see Fig.\,\ref{Fig3}.  Parameters: $\tau=0.9$, $\Delta_{\alpha}\equiv\Delta$, $\varepsilon_{\alpha_j}=0$,  $\Gamma_{\rm L_j}=0.7\Delta$, $\Gamma_{\rm R_j}=0$.}
\label{Fig5} 
\end{figure}

\section{Conductance signatures of Andreev exceptional points}
Having demonstrated the emergence of Andreev EPs in minimal Kitaev chain Josephson junctions,  we now discuss their detection by means of conductance. For this purpose, we attach two normal leads to the leftmost (L) and rightmost (R) sides of the non-Hermitian Josephson junction, denoted by $N_{\rm L,R}$ in Fig.\,\ref{Fig1}, with their couplings to the system characterized  by $\Gamma_{\rm L,R}$ [Eq.\,(\ref{SigmaTotal})]. Then, we obtain conductance by using the $S$-matrix in the wide-band limit as
\begin{equation}
\label{Smatrix}
S(\omega,\phi)=\left\{1-2iW[\omega-H_{\rm eff}(0,\phi)]^{-1}W^{\dagger}\right\}\,,
\end{equation}
where $H_{\rm ef}(0,\phi)=H_{\rm JJ}(\phi)+iW^{\dagger}W$ is the effective non-Hermitian Hamiltonian  describing our open Josephson junction given by Eq.\,(\ref{Eqeff}), while $W$ is the tunneling matrix characterizing the coupling to all the leads attached to the Josephson junction, as depicted in Fig.\,\ref{Fig1}. Note that $-iW^{\dagger}W$ is equal to the retarded self-energy given by Eq.\,(\ref{SigmaTotal}). Then, using the elements of the $S$-matrix in Eq.\,(\ref{Smatrix}), the zero-temperature differential conductance is obtained as $G_{\alpha\beta}(\omega,\phi)\equiv dI_{\alpha}/dV_{\beta}$, with 
\begin{equation}
G_{\alpha\beta}(\omega,\phi)=\frac{e^{2}}{h}[\delta_{\alpha\beta}-|S_{ee}^{\alpha\beta}(\omega,\phi)|^{2}+|S_{he}^{\alpha\beta}(\omega,\phi)|^{2}]
\end{equation}
and $\alpha,\beta=\{{\rm L,R}\}$,  with $\omega$ here playing the role of the bias voltage. We also contrast the  conductance signatures with  the spectral function, calculated as
\begin{equation}
A_{\alpha_{j}\alpha_{j}}(\omega,\phi)=-\frac{1}{\pi}{\rm Im}[\mathcal{G}_{\alpha_{j}\alpha_{j}}^{r}(\omega,\phi)]\,,
\end{equation} 
where $\mathcal{G}_{\alpha_{j}\alpha_{j}}^{r}(\omega,\phi)$ is the $\alpha_{j}\alpha_{j}$ diagonal element  of the system retarded Green's function $\mathcal{G}^{r}(\omega,\phi)=[\omega-H_{\rm ef}(0,\phi)]^{-1}$, with $\alpha=L (R)$ denoting the left (right) minimal Kitaev chain and $j=1 (2)$ denoting the first (second) QD. 

We first focus on the local conductances $G_{\rm LL}$ and $G_{\rm RR}$, together with their spectral counterparts, in the sweet-spot regime $\Delta_{\alpha}=t_{\alpha}$ and $\varepsilon_{\alpha_j}=0$. These quantities are presented in Fig.\,\ref{Fig5} as a function of $\omega$ and $\phi$; note that this sweet-spot regime corresponds to the one reported in Fig.\,\ref{Fig3}(a), where non-Hermiticity is     the same in the two QDs of the left minimal Kitaev chain. In this case, the local conductance $G_{\rm LL}$ is tiny, but exhibits noticeably   larger values along the zero Re energy line connecting the EPs, as indicated by the cyan arrow in Fig.\,\ref{Fig5}(a).   In this regard, $G_{\rm LL}$  is unable to signal the emergence of EPs at higher energies, nor does   it show  direct features of the dispersionless levels at zero energy. The difficulty in identifying the zero-energy Majorana states via $G_{\rm LL}$  is due to the large values of non-Hermiticity applied to the left minimal Kitaev chain; this gives rise to a broad resonance around zero frequency [Fig.\,\ref{Fig3}(b)]. The features revealed by the local conductance  $G_{\rm LL}$ are also revealed  by the spectral function at the leftmost QD, $A_{\rm L_{1}L_{1}}$, see Fig.\,\ref{Fig5}(a). This is, of course, expected   since local conductance measures the spectral weight at the point where the lead is attached. In fact, by inspecting the spectral function at the second QD of the left chain ($A_{\rm L_{2}L_{2}}$), we clearly see that the signature of the zero Re energy line is absent [Fig.\,\ref{Fig5}(b)]; there is, however, a brighter line  at the position of the finite Re energy EPs obtained in Fig.\,\ref{Fig3}(a), indicating that such EPs can be detected by measuring conductance at the second QD of the left chain.

When inspecting the local conductance $G_{\rm RR}$ in Fig.\,\ref{Fig5}(d), the strongest signal comes from   zero frequency, which stems from the dispersionless zero energy Majorana states located at the outer QDs and shown in Fig.\,\ref{Fig3}(a). This local conductance also shows very tiny but finite values at the positions of the EPs at higher energies, which are, however, hard to identify by the naked eye. These local conductance features obtained from the second QD in the right minimal Kitaev chain are also reflected  by the spectral function at that  QD ($A_{\rm R_{2}R_{2}}$), with clear signals due to the dispersionless zero-energy Majorana states and the finite Re energy line connecting the EPs, see  the green arrows Fig.\,\ref{Fig5}(f). Interestingly,  $A_{\rm R_{1}R_{1}}$ unveils the emergence of both zero and finite Re energy EPs, which is reflected by the strongest spectral weights at the positions of the zero and finite Re energy lines, see cyan and green arrows in Fig.\,\ref{Fig5}(e). Thus, in order to detect the zero and finite Re energy EPs, along with their connecting lines, shown in Fig.\,\ref{Fig3}(a), it is convenient to perform conductance measurements  at the first QD of the second minimal Kitaev chain in our setup [Fig.\,\ref{Fig1}]. Conversely, if the interest is to probe the dispersionless Majorana modes located at the outer QDs, it is advisable  to do   so by means of local conductance at the second QD of the right minimal Kitaev chain. We have verified that the nonlocal conductance in the sweet-spot regime acquires vanishing values, challenging its use for identifying EPs.

 \begin{figure}[!t]
\centering
\includegraphics[width=0.49\textwidth]{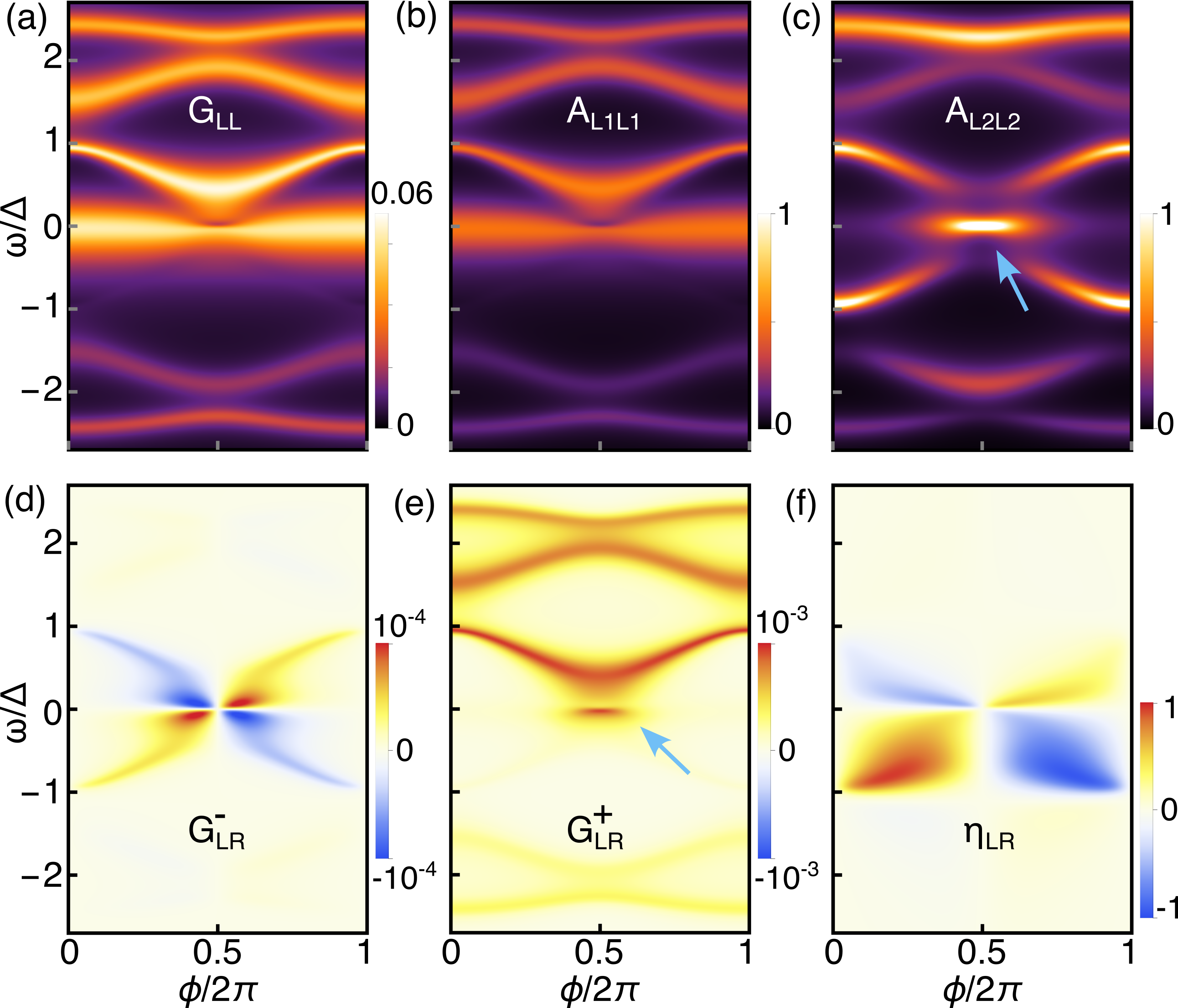}
\caption{(a-c) Local conductance $G_{\rm LL}$ (a) and spectral function $A_{\rm L_{j}L_{j}}$ (b,c) as a function of frequency $\omega$ and $\phi$ at $\Delta_{\alpha}\neq t_{\alpha}$, $\varepsilon_{\alpha_{j}}\neq0$ and $\Gamma_{\rm L_1, R_{2}}\neq0$, $\Gamma_{\rm L_2,R_1}=0$. The conductance is given in units of $e^{2}/h$. In (c),  the color scale is cut off at 1 for visualization. 
The light blue   arrow  indicates the appearance of the zero   Re energy line connecting   EPs found at the sweet spot, see Fig.\,\ref{Fig4}. (d,e) Antisymmetric ($G_{\rm LR}^{-}$) and symmetric ($G_{\rm LR}^{+}$) nonlocal conductance as a function of $\omega$ and $\phi$, where $G_{\rm LR}^{\pm}=(G_{\rm LR}\pm G_{\rm RL})/2$. (f) Antisymmetric conductance normalized to its symmetric counterpart, denoted as $\eta_{\rm LR}=G_{\rm LR}^{-}/G_{\rm LR}^{+}$. Parameters: $\tau=0.9$, $\Delta_{\alpha}\equiv\Delta$, $t_{\alpha}=0.8\Delta$, $\varepsilon_{\alpha_j}=0.5\Delta$,  $\Gamma_{\rm L_1,R_2}=0.3\Delta$, $\Gamma_{\rm L_2,R_1}=0$.}
\label{Fig6} 
\end{figure}

Away from the sweet spot, where $\Delta_{\alpha}\neq t_{\alpha}$, local and nonlocal conductance can exhibit   evidence of  the EPs shown in Fig.\,\ref{Fig4}, but the signatures need to be carefully analyzed. For instance,  when  the outer QDs are subjected to equal non-Hermiticity and $\varepsilon_{\alpha_j}=\varepsilon\neq0$, the local and nonlocal conductances ($G_{\rm LL}$ and $G_{\rm LR}$) and the spectral functions in the QDs of the left minimal Kitaev chain ($A_{\rm L_{j}L_{j}}$) show the emergence of the zero Re energy EPs presented in Fig.\,\ref{Fig4}(c); here, $t_{\alpha}=0.8\Delta_{\alpha}$.
 To demonstrate this, in  in Fig.\,\ref{Fig6} we present $G_{\rm LL}$, $G_{\rm LR}$, and $A_{\rm L_{j}L_{j}}$  as functions of $\omega$ and $\phi$. We note that $G_{\rm RR}$ and $A_{\rm R_{j}R_{j}}$ give  the same information as $G_{\rm LL}$ and $A_{\rm L_{j}L_{j}}$, respectively, which is expected since the non-Hermitian effects and onsite energies are symmetrically distributed along the Josephson junction. In this regard, the local conductance $G_{\rm LL}$ shows strong broad intensities at the energies of the ABSs, although the signals of the zero Re energy EPs are   not very clear.   These broad intensities occur because this regime considers non-Hermiticity at the outer QDs of the Josephson junction. Moreover, a closer inspection shows that the broad strong zero energy signal is not really uniform for all $\phi$ but its width (along the frequency axis) reduces along the zero Re energy line connecting the respective EPs around $\phi=\pi$. The broad, seemingly resonance outside such zero Re energy line (below and above the EPs) consists of two resonances, since there are two levels around zero Re energy in this case [Fig.\,\ref{Fig4}(c)]; however, these nearly zero Re energy levels have large Im parts, which   produce broad conductance resonances that are almost superimposed and are, hence, hard to properly resolve. The same signature is provided by the spectral function at the leftmost QD of the left chain, see  Fig.\,\ref{Fig6}(b). In contrast to the signals at the leftmost QD, accessing the spectral function in the second  QD of the left    chain offers a   clearer way to identify the ABSs and, interestingly, also reveals the emergence of the zero Re energy line connecting the EPs, see  the cyan arrow in  Fig.\,\ref{Fig6}(c). 

Besides local signals, the nonlocal conductance   is also capable of revealing features of the zero Re energy EPs and their connecting energy line. In this case, it is useful to analyze the symmetric and antisymmetric combinations of $G_{\rm LR}$ and $G_{\rm RL}$ as $G^{\pm}_{\rm LR}=(G_{\rm LR}\pm G_{\rm RL})/2$, which are presented in   Fig.\,\ref{Fig6}(d,e) as a function of $\omega$ and $\phi$. The antisymmetric combination ($G^{-}_{\rm LR}$) reveals the formation of  subgap  ABSs,  along which $G^{-}_{\rm LR}$ takes on negative values for the descending (ascending) ABS within $\omega>0$, while the opposite is true for $\omega<0$. The negative  values of $G^{-}_{\rm LR}$ for $\omega>0$  mean that $G_{\rm RL}>G_{\rm LR}$, which is, of course, tied to ECT processes being smaller than CAR ($t<\Delta$) in this regime. At $\omega\rightarrow0$,  $G^{-}_{\rm LR}$ is vanishingly small and exhibits a clear  roughly constant  minigap that runs almost over the entire length of the zero Re energy line connecting  the EPs, see  Fig.\,\ref{Fig6}(d) and Fig.\,\ref{Fig4}(c). Even more interestingly, the symmetric nonlocal  conductance ($G^{+}_{\rm LR}$) is positive for all $\omega$ and $\phi$, and shows a strong weight at the zero Re energy line connecting the EPs as well as at the positive Andreev levels, see the cyan arrow and red intensities in Fig.\,\ref{Fig6}(e).  We can also quantify the ratio between antisymmetric and symmetric conductances by $\eta_{\rm LR}=G^{-}_{\rm LR}/G^{+}_{\rm LR}$, which not only provides information about where $G_{\rm LR (RL)}$ dominates, but also about the size of $G^{\pm}$. Since the factor $\eta_{\rm LR}$ characterizes the LR and RL conductances,  its finite value indicates that the system exhibits a spatial asymmetry that is inherent to the ratio between CAR and ECT. Overall, local and nonlocal conductances, in either the sweet-spot or away from the sweet-spot regimes, can detect the formation of EPs and their connecting  lines, which requires special care about where the conductance is measured in the minimal Kitaev chain Josephson junction.

\section{Conclusions}
\label{section4}
We have investigated the impact of non-Hermiticity on Josephson junctions formed by minimal Kitaev chains. In particular, we have considered non-Hermiticity arising from coupling each quantum dot in both chains  to normal leads. The effect of the leads renders  our setup open, which we describe  by an effective non-Hermitian Hamiltonian exhibiting a complex Andreev spectrum. Interestingly, we have shown that the  complex spectrum hosts non-Hermitian degeneracies, referred to as Andreev exceptional points, at which eigenvalues and eigenvectors coalesce while being entirely controlled by the superconducting phase difference between the chains. We have found that, depending on the spatial  unequal distribution of non-Hermiticity, such Andreev exceptional points can form at zero and finite real energies, which are connected by stable energy lines and protected by   non-Hermitian topology.  Of particular relevance are the exceptional points at zero real energy because their connecting lines stabilize Majorana states over a continuous set of points rather than at a single point, as   occurs in the Hermitian regime. Moreover, we have  unveiled that the Andreev exceptional points can form along lines, here coined here Andreev exceptional lines, which enclose two-dimensional zero-energy regions in the space defined by  the superconducting phase and one extra system parameter, such as the onsite energies, the ratio between electron cotunneling and crossed Andreev reflections, or the strength of non-Hermiticity. Finally, we have demonstrated that the emergent Andreev exceptional points  and their connecting lines  can be detected by means of local and nonlocal conductances, with the latter being more efficient when electron cotunneling and crossed Andreev reflections are distinct. Our findings may be helpful for inducing broader regions with Majorana modes and put forward the realization of topological phases by combining few-site Kitaev chains, non-Hermitian topology, and the Josephson effect.

\section{Acknowledgements}
J. C. acknowledges  financial support from the Swedish Research Council  (Vetenskapsr\aa det Grant No.~2021-04121) and from the Olle Engkvist Foundation (Grant No. 243-1026). M. S. are supported by JST CREST (Grant No. JPMJCR19T2) and JSPS KAKENHI (Grant Nos. JP24K00569 and JP25H01250). The computations  were  enabled by resources provided by the National Academic Infrastructure for Supercomputing in Sweden (NAISS), partially funded by the Swedish Research Council through grant agreement no. 2022-06725.

 \section*{Data availability statement}
  The data supporting the findings of this work are available from the  authors upon reasonable request.

\bibliography{biblio}

\begin{thebibliography}{96}%
\makeatletter
\providecommand \@ifxundefined [1]{%
 \@ifx{#1\undefined}
}%
\providecommand \@ifnum [1]{%
 \ifnum #1\expandafter \@firstoftwo
 \else \expandafter \@secondoftwo
 \fi
}%
\providecommand \@ifx [1]{%
 \ifx #1\expandafter \@firstoftwo
 \else \expandafter \@secondoftwo
 \fi
}%
\providecommand \natexlab [1]{#1}%
\providecommand \enquote  [1]{``#1''}%
\providecommand \bibnamefont  [1]{#1}%
\providecommand \bibfnamefont [1]{#1}%
\providecommand \citenamefont [1]{#1}%
\providecommand \href@noop [0]{\@secondoftwo}%
\providecommand \href [0]{\begingroup \@sanitize@url \@href}%
\providecommand \@href[1]{\@@startlink{#1}\@@href}%
\providecommand \@@href[1]{\endgroup#1\@@endlink}%
\providecommand \@sanitize@url [0]{\catcode `\\12\catcode `\$12\catcode
  `\&12\catcode `\#12\catcode `\^12\catcode `\_12\catcode `\%12\relax}%
\providecommand \@@startlink[1]{}%
\providecommand \@@endlink[0]{}%
\providecommand \url  [0]{\begingroup\@sanitize@url \@url }%
\providecommand \@url [1]{\endgroup\@href {#1}{\urlprefix }}%
\providecommand \urlprefix  [0]{URL }%
\providecommand \Eprint [0]{\href }%
\providecommand \doibase [0]{https://doi.org/}%
\providecommand \selectlanguage [0]{\@gobble}%
\providecommand \bibinfo  [0]{\@secondoftwo}%
\providecommand \bibfield  [0]{\@secondoftwo}%
\providecommand \translation [1]{[#1]}%
\providecommand \BibitemOpen [0]{}%
\providecommand \bibitemStop [0]{}%
\providecommand \bibitemNoStop [0]{.\EOS\space}%
\providecommand \EOS [0]{\spacefactor3000\relax}%
\providecommand \BibitemShut  [1]{\csname bibitem#1\endcsname}%
\let\auto@bib@innerbib\@empty
\bibitem [{\citenamefont {Dvir}\ \emph {et~al.}(2023)\citenamefont {Dvir},
  \citenamefont {Wang}, \citenamefont {van Loo}, \citenamefont {Liu},
  \citenamefont {Mazur}, \citenamefont {Bordin}, \citenamefont {Ten~Haaf},
  \citenamefont {Wang}, \citenamefont {van Driel}, \citenamefont {Zatelli}
  \emph {et~al.}}]{dvir2023realization}%
  \BibitemOpen
  \bibfield  {author} {\bibinfo {author} {\bibfnamefont {T.}~\bibnamefont
  {Dvir}}, \bibinfo {author} {\bibfnamefont {G.}~\bibnamefont {Wang}}, \bibinfo
  {author} {\bibfnamefont {N.}~\bibnamefont {van Loo}}, \bibinfo {author}
  {\bibfnamefont {C.-X.}\ \bibnamefont {Liu}}, \bibinfo {author} {\bibfnamefont
  {G.~P.}\ \bibnamefont {Mazur}}, \bibinfo {author} {\bibfnamefont
  {A.}~\bibnamefont {Bordin}}, \bibinfo {author} {\bibfnamefont {S.~L.}\
  \bibnamefont {Ten~Haaf}}, \bibinfo {author} {\bibfnamefont {J.-Y.}\
  \bibnamefont {Wang}}, \bibinfo {author} {\bibfnamefont {D.}~\bibnamefont {van
  Driel}}, \bibinfo {author} {\bibfnamefont {F.}~\bibnamefont {Zatelli}}, \emph
  {et~al.},\ }\bibfield  {title} {\bibinfo {title} {Realization of a minimal
  {K}itaev chain in coupled quantum dots},\ }\href
  {https://doi.org/10.1038/s41586-022-05585-1} {\bibfield  {journal} {\bibinfo
  {journal} {Nature}\ }\textbf {\bibinfo {volume} {614}},\ \bibinfo {pages}
  {445} (\bibinfo {year} {2023})}\BibitemShut {NoStop}%
\bibitem [{\citenamefont {ten Haaf}\ \emph {et~al.}(2024)\citenamefont {ten
  Haaf}, \citenamefont {Wang}, \citenamefont {Bozkurt}, \citenamefont {Liu},
  \citenamefont {Kulesh}, \citenamefont {Kim}, \citenamefont {Xiao},
  \citenamefont {Thomas}, \citenamefont {Manfra}, \citenamefont {Dvir},
  \citenamefont {Wimmer},\ and\ \citenamefont {Goswami}}]{Haaf2024}%
  \BibitemOpen
  \bibfield  {author} {\bibinfo {author} {\bibfnamefont {S.~L.~D.}\
  \bibnamefont {ten Haaf}}, \bibinfo {author} {\bibfnamefont {Q.}~\bibnamefont
  {Wang}}, \bibinfo {author} {\bibfnamefont {A.~M.}\ \bibnamefont {Bozkurt}},
  \bibinfo {author} {\bibfnamefont {C.-X.}\ \bibnamefont {Liu}}, \bibinfo
  {author} {\bibfnamefont {I.}~\bibnamefont {Kulesh}}, \bibinfo {author}
  {\bibfnamefont {P.}~\bibnamefont {Kim}}, \bibinfo {author} {\bibfnamefont
  {D.}~\bibnamefont {Xiao}}, \bibinfo {author} {\bibfnamefont {C.}~\bibnamefont
  {Thomas}}, \bibinfo {author} {\bibfnamefont {M.~J.}\ \bibnamefont {Manfra}},
  \bibinfo {author} {\bibfnamefont {T.}~\bibnamefont {Dvir}}, \bibinfo {author}
  {\bibfnamefont {M.}~\bibnamefont {Wimmer}},\ and\ \bibinfo {author}
  {\bibfnamefont {S.}~\bibnamefont {Goswami}},\ }\bibfield  {title} {\bibinfo
  {title} {A two-site {K}itaev chain in a two-dimensional electron gas},\
  }\href {https://doi.org/10.1038/s41586-024-07434-9} {\bibfield  {journal}
  {\bibinfo  {journal} {Nature}\ }\textbf {\bibinfo {volume} {630}},\ \bibinfo
  {pages} {329} (\bibinfo {year} {2024})}\BibitemShut {NoStop}%
\bibitem [{\citenamefont {Zatelli}\ \emph {et~al.}(2024)\citenamefont
  {Zatelli}, \citenamefont {van Driel}, \citenamefont {Xu}, \citenamefont
  {Wang}, \citenamefont {Liu}, \citenamefont {Bordin}, \citenamefont {Roovers},
  \citenamefont {Mazur}, \citenamefont {van Loo}, \citenamefont {Wolff},
  \citenamefont {Bozkurt}, \citenamefont {Badawy}, \citenamefont {Gazibegovic},
  \citenamefont {Bakkers}, \citenamefont {Wimmer}, \citenamefont
  {Kouwenhoven},\ and\ \citenamefont {Dvir}}]{Zatelli_2024}%
  \BibitemOpen
  \bibfield  {author} {\bibinfo {author} {\bibfnamefont {F.}~\bibnamefont
  {Zatelli}}, \bibinfo {author} {\bibfnamefont {D.}~\bibnamefont {van Driel}},
  \bibinfo {author} {\bibfnamefont {D.}~\bibnamefont {Xu}}, \bibinfo {author}
  {\bibfnamefont {G.}~\bibnamefont {Wang}}, \bibinfo {author} {\bibfnamefont
  {C.-X.}\ \bibnamefont {Liu}}, \bibinfo {author} {\bibfnamefont
  {A.}~\bibnamefont {Bordin}}, \bibinfo {author} {\bibfnamefont
  {B.}~\bibnamefont {Roovers}}, \bibinfo {author} {\bibfnamefont {G.~P.}\
  \bibnamefont {Mazur}}, \bibinfo {author} {\bibfnamefont {N.}~\bibnamefont
  {van Loo}}, \bibinfo {author} {\bibfnamefont {J.~C.}\ \bibnamefont {Wolff}},
  \bibinfo {author} {\bibfnamefont {A.~M.}\ \bibnamefont {Bozkurt}}, \bibinfo
  {author} {\bibfnamefont {G.}~\bibnamefont {Badawy}}, \bibinfo {author}
  {\bibfnamefont {S.}~\bibnamefont {Gazibegovic}}, \bibinfo {author}
  {\bibfnamefont {E.~P. A.~M.}\ \bibnamefont {Bakkers}}, \bibinfo {author}
  {\bibfnamefont {M.}~\bibnamefont {Wimmer}}, \bibinfo {author} {\bibfnamefont
  {L.~P.}\ \bibnamefont {Kouwenhoven}},\ and\ \bibinfo {author} {\bibfnamefont
  {T.}~\bibnamefont {Dvir}},\ }\bibfield  {title} {\bibinfo {title} {Robust
  poor man’s majorana zero modes using yu-shiba-rusinov states},\ }\href
  {http://dx.doi.org/10.1038/s41467-024-52066-2} {\bibfield  {journal}
  {\bibinfo  {journal} {Nat. Commun.}\ }\textbf {\bibinfo {volume} {15}},\
  \bibinfo {pages} {7933} (\bibinfo {year} {2024})}\BibitemShut {NoStop}%
\bibitem [{\citenamefont {Bordin}\ \emph {et~al.}(2025)\citenamefont {Bordin},
  \citenamefont {Liu}, \citenamefont {Dvir}, \citenamefont {Zatelli},
  \citenamefont {Ten~Haaf}, \citenamefont {van Driel}, \citenamefont {Wang},
  \citenamefont {Van~Loo}, \citenamefont {Zhang}, \citenamefont {Wolff} \emph
  {et~al.}}]{bordin2025enhanced}%
  \BibitemOpen
  \bibfield  {author} {\bibinfo {author} {\bibfnamefont {A.}~\bibnamefont
  {Bordin}}, \bibinfo {author} {\bibfnamefont {C.-X.}\ \bibnamefont {Liu}},
  \bibinfo {author} {\bibfnamefont {T.}~\bibnamefont {Dvir}}, \bibinfo {author}
  {\bibfnamefont {F.}~\bibnamefont {Zatelli}}, \bibinfo {author} {\bibfnamefont
  {S.~L.}\ \bibnamefont {Ten~Haaf}}, \bibinfo {author} {\bibfnamefont
  {D.}~\bibnamefont {van Driel}}, \bibinfo {author} {\bibfnamefont
  {G.}~\bibnamefont {Wang}}, \bibinfo {author} {\bibfnamefont {N.}~\bibnamefont
  {Van~Loo}}, \bibinfo {author} {\bibfnamefont {Y.}~\bibnamefont {Zhang}},
  \bibinfo {author} {\bibfnamefont {J.~C.}\ \bibnamefont {Wolff}}, \emph
  {et~al.},\ }\bibfield  {title} {\bibinfo {title} {Enhanced majorana stability
  in a three-site kitaev chain},\ }\href
  {https://doi.org/10.1038/s41565-025-01894-4} {\bibfield  {journal} {\bibinfo
  {journal} {Nat. Nanotech.}\ }\textbf {\bibinfo {volume} {20}},\ \bibinfo
  {pages} {726} (\bibinfo {year} {2025})}\BibitemShut {NoStop}%
\bibitem [{\citenamefont {Ten~Haaf}\ \emph {et~al.}(2025)\citenamefont
  {Ten~Haaf}, \citenamefont {Zhang}, \citenamefont {Wang}, \citenamefont
  {Bordin}, \citenamefont {Liu}, \citenamefont {Kulesh}, \citenamefont
  {Sietses}, \citenamefont {Prosko}, \citenamefont {Xiao}, \citenamefont
  {Thomas}, \citenamefont {Manfra}, \citenamefont {Wimmer},\ and\ \citenamefont
  {Goswami}}]{ten2025observation}%
  \BibitemOpen
  \bibfield  {author} {\bibinfo {author} {\bibfnamefont {S.~L.~D.}\
  \bibnamefont {Ten~Haaf}}, \bibinfo {author} {\bibfnamefont {Y.}~\bibnamefont
  {Zhang}}, \bibinfo {author} {\bibfnamefont {Q.}~\bibnamefont {Wang}},
  \bibinfo {author} {\bibfnamefont {A.}~\bibnamefont {Bordin}}, \bibinfo
  {author} {\bibfnamefont {C.-X.}\ \bibnamefont {Liu}}, \bibinfo {author}
  {\bibfnamefont {I.}~\bibnamefont {Kulesh}}, \bibinfo {author} {\bibfnamefont
  {V.~P.~M.}\ \bibnamefont {Sietses}}, \bibinfo {author} {\bibfnamefont
  {C.~G.}\ \bibnamefont {Prosko}}, \bibinfo {author} {\bibfnamefont
  {D.}~\bibnamefont {Xiao}}, \bibinfo {author} {\bibfnamefont {C.}~\bibnamefont
  {Thomas}}, \bibinfo {author} {\bibfnamefont {J.~M.}\ \bibnamefont {Manfra}},
  \bibinfo {author} {\bibfnamefont {M.}~\bibnamefont {Wimmer}},\ and\ \bibinfo
  {author} {\bibfnamefont {S.}~\bibnamefont {Goswami}},\ }\bibfield  {title}
  {\bibinfo {title} {Observation of edge and bulk states in a three-site kitaev
  chain},\ }\href {https://doi.org/10.1038/s41586-025-08892-5} {\bibfield
  {journal} {\bibinfo  {journal} {Nature}\ }\textbf {\bibinfo {volume} {641}},\
  \bibinfo {pages} {890} (\bibinfo {year} {2025})}\BibitemShut {NoStop}%
\bibitem [{\citenamefont {Bordin}\ \emph {et~al.}(2026)\citenamefont {Bordin},
  \citenamefont {Bennebroek~Evertsz’}, \citenamefont {Roovers}, \citenamefont
  {Torres~Luna}, \citenamefont {Huisman}, \citenamefont {Zatelli},
  \citenamefont {Mazur}, \citenamefont {Ten~Haaf}, \citenamefont {Badawy},
  \citenamefont {Bakkers}, \citenamefont {Liu}, \citenamefont {Souto},
  \citenamefont {van Loo},\ and\ \citenamefont
  {Kouwenhoven}}]{bordin2026probing}%
  \BibitemOpen
  \bibfield  {author} {\bibinfo {author} {\bibfnamefont {A.}~\bibnamefont
  {Bordin}}, \bibinfo {author} {\bibfnamefont {F.~J.}\ \bibnamefont
  {Bennebroek~Evertsz’}}, \bibinfo {author} {\bibfnamefont {B.}~\bibnamefont
  {Roovers}}, \bibinfo {author} {\bibfnamefont {J.~D.}\ \bibnamefont
  {Torres~Luna}}, \bibinfo {author} {\bibfnamefont {W.~D.}\ \bibnamefont
  {Huisman}}, \bibinfo {author} {\bibfnamefont {F.}~\bibnamefont {Zatelli}},
  \bibinfo {author} {\bibfnamefont {G.~P.}\ \bibnamefont {Mazur}}, \bibinfo
  {author} {\bibfnamefont {S.~L.~D.}\ \bibnamefont {Ten~Haaf}}, \bibinfo
  {author} {\bibfnamefont {G.}~\bibnamefont {Badawy}}, \bibinfo {author}
  {\bibfnamefont {E.~P. A.~M.}\ \bibnamefont {Bakkers}}, \bibinfo {author}
  {\bibfnamefont {C.-X.}\ \bibnamefont {Liu}}, \bibinfo {author} {\bibfnamefont
  {R.~S.}\ \bibnamefont {Souto}}, \bibinfo {author} {\bibfnamefont
  {N.}~\bibnamefont {van Loo}},\ and\ \bibinfo {author} {\bibfnamefont {L.~P.}\
  \bibnamefont {Kouwenhoven}},\ }\bibfield  {title} {\bibinfo {title} {Probing
  {M}ajorana localization of a phase-controlled three-site {K}itaev chain with
  an additional quantum dot},\ }\href
  {https://doi.org/10.1038/s41467-026-68897-0} {\bibfield  {journal} {\bibinfo
  {journal} {Nat. Commun.}\ }\textbf {\bibinfo {volume} {17}},\ \bibinfo
  {pages} {2313} (\bibinfo {year} {2026})}\BibitemShut {NoStop}%
\bibitem [{\citenamefont {van Loo}\ \emph {et~al.}(2026)\citenamefont {van
  Loo}, \citenamefont {Zatelli}, \citenamefont {Steffensen}, \citenamefont
  {Roovers}, \citenamefont {Wang}, \citenamefont {Van~Caekenberghe},
  \citenamefont {Bordin}, \citenamefont {van Driel}, \citenamefont {Zhang},
  \citenamefont {Huisman}, \citenamefont {Badawy}, \citenamefont {Bakkers},
  \citenamefont {Mazur}, \citenamefont {Aguado},\ and\ \citenamefont
  {Kouwenhoven}}]{van2026single}%
  \BibitemOpen
  \bibfield  {author} {\bibinfo {author} {\bibfnamefont {N.}~\bibnamefont {van
  Loo}}, \bibinfo {author} {\bibfnamefont {F.}~\bibnamefont {Zatelli}},
  \bibinfo {author} {\bibfnamefont {G.~O.}\ \bibnamefont {Steffensen}},
  \bibinfo {author} {\bibfnamefont {B.}~\bibnamefont {Roovers}}, \bibinfo
  {author} {\bibfnamefont {G.}~\bibnamefont {Wang}}, \bibinfo {author}
  {\bibfnamefont {T.}~\bibnamefont {Van~Caekenberghe}}, \bibinfo {author}
  {\bibfnamefont {A.}~\bibnamefont {Bordin}}, \bibinfo {author} {\bibfnamefont
  {D.}~\bibnamefont {van Driel}}, \bibinfo {author} {\bibfnamefont
  {Y.}~\bibnamefont {Zhang}}, \bibinfo {author} {\bibfnamefont {W.~D.}\
  \bibnamefont {Huisman}}, \bibinfo {author} {\bibfnamefont {G.}~\bibnamefont
  {Badawy}}, \bibinfo {author} {\bibfnamefont {E.~P. A.~M.}\ \bibnamefont
  {Bakkers}}, \bibinfo {author} {\bibfnamefont {G.~P.}\ \bibnamefont {Mazur}},
  \bibinfo {author} {\bibfnamefont {R.}~\bibnamefont {Aguado}},\ and\ \bibinfo
  {author} {\bibfnamefont {L.~P.}\ \bibnamefont {Kouwenhoven}},\ }\bibfield
  {title} {\bibinfo {title} {Single-shot parity readout of a minimal {K}itaev
  chain},\ }\href {https://doi.org/10.1038/s41586-025-09927-7} {\bibfield
  {journal} {\bibinfo  {journal} {Nature}\ }\textbf {\bibinfo {volume} {650}},\
  \bibinfo {pages} {334} (\bibinfo {year} {2026})}\BibitemShut {NoStop}%
\bibitem [{\citenamefont {van Driel}\ \emph {et~al.}(2026)\citenamefont {van
  Driel}, \citenamefont {Koch}, \citenamefont {Sietses}, \citenamefont {Haaf},
  \citenamefont {Liu}, \citenamefont {Zatelli}, \citenamefont {Roovers},
  \citenamefont {Bordin}, \citenamefont {van Loo}, \citenamefont {Wang},
  \citenamefont {Wolff}, \citenamefont {Mazur}, \citenamefont {Dvir},
  \citenamefont {Kulesh}, \citenamefont {Wang}, \citenamefont {Bozkurt},
  \citenamefont {Gazibegovic}, \citenamefont {Badawy}, \citenamefont {Bakkers},
  \citenamefont {Wimmer}, \citenamefont {Goswami}, \citenamefont {Lado},
  \citenamefont {Kouwenhoven},\ and\ \citenamefont {Greplova}}]{ryll-qb42}%
  \BibitemOpen
  \bibfield  {author} {\bibinfo {author} {\bibfnamefont {D.}~\bibnamefont {van
  Driel}}, \bibinfo {author} {\bibfnamefont {R.}~\bibnamefont {Koch}}, \bibinfo
  {author} {\bibfnamefont {V.~P.~M.}\ \bibnamefont {Sietses}}, \bibinfo
  {author} {\bibfnamefont {S.~L. D.~t.}\ \bibnamefont {Haaf}}, \bibinfo
  {author} {\bibfnamefont {C.-X.}\ \bibnamefont {Liu}}, \bibinfo {author}
  {\bibfnamefont {F.}~\bibnamefont {Zatelli}}, \bibinfo {author} {\bibfnamefont
  {B.}~\bibnamefont {Roovers}}, \bibinfo {author} {\bibfnamefont
  {A.}~\bibnamefont {Bordin}}, \bibinfo {author} {\bibfnamefont
  {N.}~\bibnamefont {van Loo}}, \bibinfo {author} {\bibfnamefont
  {G.}~\bibnamefont {Wang}}, \bibinfo {author} {\bibfnamefont {J.~C.}\
  \bibnamefont {Wolff}}, \bibinfo {author} {\bibfnamefont {G.~P.}\ \bibnamefont
  {Mazur}}, \bibinfo {author} {\bibfnamefont {T.}~\bibnamefont {Dvir}},
  \bibinfo {author} {\bibfnamefont {I.}~\bibnamefont {Kulesh}}, \bibinfo
  {author} {\bibfnamefont {Q.}~\bibnamefont {Wang}}, \bibinfo {author}
  {\bibfnamefont {A.~M.}\ \bibnamefont {Bozkurt}}, \bibinfo {author}
  {\bibfnamefont {S.}~\bibnamefont {Gazibegovic}}, \bibinfo {author}
  {\bibfnamefont {G.}~\bibnamefont {Badawy}}, \bibinfo {author} {\bibfnamefont
  {E.~P. A.~M.}\ \bibnamefont {Bakkers}}, \bibinfo {author} {\bibfnamefont
  {M.}~\bibnamefont {Wimmer}}, \bibinfo {author} {\bibfnamefont
  {S.}~\bibnamefont {Goswami}}, \bibinfo {author} {\bibfnamefont {J.~L.}\
  \bibnamefont {Lado}}, \bibinfo {author} {\bibfnamefont {L.~P.}\ \bibnamefont
  {Kouwenhoven}},\ and\ \bibinfo {author} {\bibfnamefont {E.}~\bibnamefont
  {Greplova}},\ }\bibfield  {title} {\bibinfo {title} {Cross-platform
  autonomous control of minimal {K}itaev chains},\ }\href
  {https://doi.org/10.1103/ryll-qb42} {\bibfield  {journal} {\bibinfo
  {journal} {PRX Intelligence}\ }\textbf {\bibinfo {volume} {1}},\ \bibinfo
  {pages} {013005} (\bibinfo {year} {2026})}\BibitemShut {NoStop}%
\bibitem [{\citenamefont {Zatelli}\ \emph {et~al.}(2026)\citenamefont
  {Zatelli}, \citenamefont {Roovers}, \citenamefont {van Loo}, \citenamefont
  {Lombardi}, \citenamefont {Luna}, \citenamefont {Miles}, \citenamefont
  {Sietses}, \citenamefont {Evertsz'}, \citenamefont {{n}a}, \citenamefont
  {Bordin}, \citenamefont {Badawy}, \citenamefont {Bakkers}, \citenamefont
  {Wimmer},\ and\ \citenamefont {Kouwenhoven}}]{zatelli2026MZM}%
  \BibitemOpen
  \bibfield  {author} {\bibinfo {author} {\bibfnamefont {F.}~\bibnamefont
  {Zatelli}}, \bibinfo {author} {\bibfnamefont {B.}~\bibnamefont {Roovers}},
  \bibinfo {author} {\bibfnamefont {N.}~\bibnamefont {van Loo}}, \bibinfo
  {author} {\bibfnamefont {A.}~\bibnamefont {Lombardi}}, \bibinfo {author}
  {\bibfnamefont {J.~D.~T.}\ \bibnamefont {Luna}}, \bibinfo {author}
  {\bibfnamefont {S.}~\bibnamefont {Miles}}, \bibinfo {author} {\bibfnamefont
  {V.~P.~M.}\ \bibnamefont {Sietses}}, \bibinfo {author} {\bibfnamefont
  {F.~J.~B.}\ \bibnamefont {Evertsz'}}, \bibinfo {author} {\bibfnamefont
  {P.~C.~F.}\ \bibnamefont {{n}a}}, \bibinfo {author} {\bibfnamefont
  {A.}~\bibnamefont {Bordin}}, \bibinfo {author} {\bibfnamefont
  {G.}~\bibnamefont {Badawy}}, \bibinfo {author} {\bibfnamefont {E.~P. A.~M.}\
  \bibnamefont {Bakkers}}, \bibinfo {author} {\bibfnamefont {M.}~\bibnamefont
  {Wimmer}},\ and\ \bibinfo {author} {\bibfnamefont {L.~P.}\ \bibnamefont
  {Kouwenhoven}},\ }\bibfield  {title} {\bibinfo {title} {Majorana parity qubit
  in coupled minimal {K}itaev chains},\ }\href
  {https://arxiv.org/abs/2607.09511} {\bibfield  {journal} {\bibinfo  {journal}
  {arXiv:2607.09511}\ } (\bibinfo {year} {2026})}\BibitemShut {NoStop}%
\bibitem [{\citenamefont {Leijnse}\ and\ \citenamefont
  {Flensberg}(2012)}]{Leijnse2012}%
  \BibitemOpen
  \bibfield  {author} {\bibinfo {author} {\bibfnamefont {M.}~\bibnamefont
  {Leijnse}}\ and\ \bibinfo {author} {\bibfnamefont {K.}~\bibnamefont
  {Flensberg}},\ }\bibfield  {title} {\bibinfo {title} {Parity qubits and poor
  man's {M}ajorana bound states in double quantum dots},\ }\href
  {https://doi.org/10.1103/PhysRevB.86.134528} {\bibfield  {journal} {\bibinfo
  {journal} {Phys. Rev. B}\ }\textbf {\bibinfo {volume} {86}},\ \bibinfo
  {pages} {134528} (\bibinfo {year} {2012})}\BibitemShut {NoStop}%
\bibitem [{\citenamefont {Sau}\ and\ \citenamefont {Sarma}(2012)}]{Sau2012}%
  \BibitemOpen
  \bibfield  {author} {\bibinfo {author} {\bibfnamefont {J.~D.}\ \bibnamefont
  {Sau}}\ and\ \bibinfo {author} {\bibfnamefont {S.~D.}\ \bibnamefont
  {Sarma}},\ }\bibfield  {title} {\bibinfo {title} {Realizing a robust
  practical {M}ajorana chain in a quantum-dot-superconductor linear array},\
  }\href {https://doi.org/10.1038/ncomms1966} {\bibfield  {journal} {\bibinfo
  {journal} {Nat. Commun.}\ }\textbf {\bibinfo {volume} {3}},\ \bibinfo {pages}
  {964} (\bibinfo {year} {2012})}\BibitemShut {NoStop}%
\bibitem [{\citenamefont {Sothmann}\ \emph {et~al.}(2014)\citenamefont
  {Sothmann}, \citenamefont {Weiss}, \citenamefont {Governale},\ and\
  \citenamefont {K\"onig}}]{PhysRevB.90.220501}%
  \BibitemOpen
  \bibfield  {author} {\bibinfo {author} {\bibfnamefont {B.}~\bibnamefont
  {Sothmann}}, \bibinfo {author} {\bibfnamefont {S.}~\bibnamefont {Weiss}},
  \bibinfo {author} {\bibfnamefont {M.}~\bibnamefont {Governale}},\ and\
  \bibinfo {author} {\bibfnamefont {J.}~\bibnamefont {K\"onig}},\ }\bibfield
  {title} {\bibinfo {title} {Unconventional superconductivity in double quantum
  dots},\ }\href {https://doi.org/10.1103/PhysRevB.90.220501} {\bibfield
  {journal} {\bibinfo  {journal} {Phys. Rev. B}\ }\textbf {\bibinfo {volume}
  {90}},\ \bibinfo {pages} {220501} (\bibinfo {year} {2014})}\BibitemShut
  {NoStop}%
\bibitem [{\citenamefont {Tanaka}\ \emph {et~al.}(2011)\citenamefont {Tanaka},
  \citenamefont {Sato},\ and\ \citenamefont {Nagaosa}}]{tanaka2011symmetry}%
  \BibitemOpen
  \bibfield  {author} {\bibinfo {author} {\bibfnamefont {Y.}~\bibnamefont
  {Tanaka}}, \bibinfo {author} {\bibfnamefont {M.}~\bibnamefont {Sato}},\ and\
  \bibinfo {author} {\bibfnamefont {N.}~\bibnamefont {Nagaosa}},\ }\bibfield
  {title} {\bibinfo {title} {Symmetry and topology in
  superconductors--odd-frequency pairing and edge states--},\ }\href
  {https://doi.org/10.1143/JPSJ.81.011013} {\bibfield  {journal} {\bibinfo
  {journal} {J. Phys. Soc. Jpn.}\ }\textbf {\bibinfo {volume} {81}},\ \bibinfo
  {pages} {011013} (\bibinfo {year} {2011})}\BibitemShut {NoStop}%
\bibitem [{\citenamefont {Sato}\ and\ \citenamefont
  {Fujimoto}(2016)}]{sato2016majorana}%
  \BibitemOpen
  \bibfield  {author} {\bibinfo {author} {\bibfnamefont {M.}~\bibnamefont
  {Sato}}\ and\ \bibinfo {author} {\bibfnamefont {S.}~\bibnamefont
  {Fujimoto}},\ }\bibfield  {title} {\bibinfo {title} {Majorana fermions and
  topology in superconductors},\ }\href
  {https://doi.org/10.7566/JPSJ.85.072001} {\bibfield  {journal} {\bibinfo
  {journal} {J. Phys. Soc. Jpn.}\ }\textbf {\bibinfo {volume} {85}},\ \bibinfo
  {pages} {072001} (\bibinfo {year} {2016})}\BibitemShut {NoStop}%
\bibitem [{\citenamefont {Sato}\ and\ \citenamefont
  {Ando}(2017)}]{sato2017topological}%
  \BibitemOpen
  \bibfield  {author} {\bibinfo {author} {\bibfnamefont {M.}~\bibnamefont
  {Sato}}\ and\ \bibinfo {author} {\bibfnamefont {Y.}~\bibnamefont {Ando}},\
  }\bibfield  {title} {\bibinfo {title} {Topological superconductors: a
  review},\ }\href
  {https://iopscience.iop.org/article/10.1088/1361-6633/aa6ac7/meta} {\bibfield
   {journal} {\bibinfo  {journal} {Rep. Prog. Phys.}\ }\textbf {\bibinfo
  {volume} {80}},\ \bibinfo {pages} {076501} (\bibinfo {year}
  {2017})}\BibitemShut {NoStop}%
\bibitem [{\citenamefont {Aguado}(2017)}]{Aguadoreview17}%
  \BibitemOpen
  \bibfield  {author} {\bibinfo {author} {\bibfnamefont {R.}~\bibnamefont
  {Aguado}},\ }\bibfield  {title} {\bibinfo {title} {Majorana quasiparticles in
  condensed matter},\ }\href@noop {} {\bibfield  {journal} {\bibinfo  {journal}
  {Riv. Nuovo Cimento}\ }\textbf {\bibinfo {volume} {40}},\ \bibinfo {pages}
  {523} (\bibinfo {year} {2017})}\BibitemShut {NoStop}%
\bibitem [{\citenamefont {Lutchyn}\ \emph {et~al.}(2018)\citenamefont
  {Lutchyn}, \citenamefont {Bakkers}, \citenamefont {Kouwenhoven},
  \citenamefont {Krogstrup}, \citenamefont {Marcus},\ and\ \citenamefont
  {Oreg}}]{lutchyn2018majorana}%
  \BibitemOpen
  \bibfield  {author} {\bibinfo {author} {\bibfnamefont {R.~M.}\ \bibnamefont
  {Lutchyn}}, \bibinfo {author} {\bibfnamefont {E.~P.}\ \bibnamefont
  {Bakkers}}, \bibinfo {author} {\bibfnamefont {L.~P.}\ \bibnamefont
  {Kouwenhoven}}, \bibinfo {author} {\bibfnamefont {P.}~\bibnamefont
  {Krogstrup}}, \bibinfo {author} {\bibfnamefont {C.~M.}\ \bibnamefont
  {Marcus}},\ and\ \bibinfo {author} {\bibfnamefont {Y.}~\bibnamefont {Oreg}},\
  }\bibfield  {title} {\bibinfo {title} {Majorana zero modes in
  superconductor--semiconductor heterostructures},\ }\href
  {https://doi.org/10.1038/s41578-018-0003-1} {\bibfield  {journal} {\bibinfo
  {journal} {Nat. Rev. Mater.}\ }\textbf {\bibinfo {volume} {3}},\ \bibinfo
  {pages} {52} (\bibinfo {year} {2018})}\BibitemShut {NoStop}%
\bibitem [{\citenamefont {Cayao}\ \emph {et~al.}(2020)\citenamefont {Cayao},
  \citenamefont {Triola},\ and\ \citenamefont
  {Black-Schaffer}}]{Cayao2020Oddfrequency}%
  \BibitemOpen
  \bibfield  {author} {\bibinfo {author} {\bibfnamefont {J.}~\bibnamefont
  {Cayao}}, \bibinfo {author} {\bibfnamefont {C.}~\bibnamefont {Triola}},\ and\
  \bibinfo {author} {\bibfnamefont {A.~M.}\ \bibnamefont {Black-Schaffer}},\
  }\bibfield  {title} {\bibinfo {title} {Odd-frequency superconducting pairing
  in one-dimensional systems},\ }\href
  {https://doi.org/10.1140/epjst/e2019-900168-0} {\bibfield  {journal}
  {\bibinfo  {journal} {Eur. Phys. J. Spec. Top.}\ }\textbf {\bibinfo {volume}
  {229}},\ \bibinfo {pages} {545–575} (\bibinfo {year} {2020})}\BibitemShut
  {NoStop}%
\bibitem [{\citenamefont {Flensberg}\ \emph {et~al.}(2021)\citenamefont
  {Flensberg}, \citenamefont {von Oppen},\ and\ \citenamefont
  {Stern}}]{flensberg2021engineered}%
  \BibitemOpen
  \bibfield  {author} {\bibinfo {author} {\bibfnamefont {K.}~\bibnamefont
  {Flensberg}}, \bibinfo {author} {\bibfnamefont {F.}~\bibnamefont {von
  Oppen}},\ and\ \bibinfo {author} {\bibfnamefont {A.}~\bibnamefont {Stern}},\
  }\bibfield  {title} {\bibinfo {title} {Engineered platforms for topological
  superconductivity and {M}ajorana zero modes},\ }\href
  {https://www.nature.com/articles/s41578-021-00336-6} {\bibfield  {journal}
  {\bibinfo  {journal} {Nat. Rev. Mater.}\ }\textbf {\bibinfo {volume} {6}},\
  \bibinfo {pages} {944} (\bibinfo {year} {2021})}\BibitemShut {NoStop}%
\bibitem [{\citenamefont {Tanaka}\ \emph {et~al.}(2024)\citenamefont {Tanaka},
  \citenamefont {Tamura},\ and\ \citenamefont {Cayao}}]{tanaka2024theory}%
  \BibitemOpen
  \bibfield  {author} {\bibinfo {author} {\bibfnamefont {Y.}~\bibnamefont
  {Tanaka}}, \bibinfo {author} {\bibfnamefont {S.}~\bibnamefont {Tamura}},\
  and\ \bibinfo {author} {\bibfnamefont {J.}~\bibnamefont {Cayao}},\ }\bibfield
   {title} {\bibinfo {title} {Theory of {M}ajorana zero modes in unconventional
  superconductors},\ }\href {https://doi.org/10.1093/ptep/ptae065} {\bibfield
  {journal} {\bibinfo  {journal} {Prog. Theor. Exp. Phys.}\ }\textbf {\bibinfo
  {volume} {2024}},\ \bibinfo {pages} {08C105} (\bibinfo {year}
  {2024})}\BibitemShut {NoStop}%
\bibitem [{\citenamefont {Cayao}(2024{\natexlab{a}})}]{PhysRevB.110.125408}%
  \BibitemOpen
  \bibfield  {author} {\bibinfo {author} {\bibfnamefont {J.}~\bibnamefont
  {Cayao}},\ }\bibfield  {title} {\bibinfo {title} {Emergent pair symmetries in
  systems with poor man's majorana modes},\ }\href
  {https://doi.org/10.1103/PhysRevB.110.125408} {\bibfield  {journal} {\bibinfo
   {journal} {Phys. Rev. B}\ }\textbf {\bibinfo {volume} {110}},\ \bibinfo
  {pages} {125408} (\bibinfo {year} {2024}{\natexlab{a}})}\BibitemShut
  {NoStop}%
\bibitem [{\citenamefont {Tsintzis}\ \emph {et~al.}(2022)\citenamefont
  {Tsintzis}, \citenamefont {Souto},\ and\ \citenamefont
  {Leijnse}}]{PhysRevB.106.L201404}%
  \BibitemOpen
  \bibfield  {author} {\bibinfo {author} {\bibfnamefont {A.}~\bibnamefont
  {Tsintzis}}, \bibinfo {author} {\bibfnamefont {R.~S.}\ \bibnamefont
  {Souto}},\ and\ \bibinfo {author} {\bibfnamefont {M.}~\bibnamefont
  {Leijnse}},\ }\bibfield  {title} {\bibinfo {title} {Creating and detecting
  poor man's majorana bound states in interacting quantum dots},\ }\href
  {https://doi.org/10.1103/PhysRevB.106.L201404} {\bibfield  {journal}
  {\bibinfo  {journal} {Phys. Rev. B}\ }\textbf {\bibinfo {volume} {106}},\
  \bibinfo {pages} {L201404} (\bibinfo {year} {2022})}\BibitemShut {NoStop}%
\bibitem [{\citenamefont {Seoane~Souto}\ \emph {et~al.}(2023)\citenamefont
  {Seoane~Souto}, \citenamefont {Tsintzis}, \citenamefont {Leijnse},\ and\
  \citenamefont {Danon}}]{PhysRevResearch.5.043182}%
  \BibitemOpen
  \bibfield  {author} {\bibinfo {author} {\bibfnamefont {R.}~\bibnamefont
  {Seoane~Souto}}, \bibinfo {author} {\bibfnamefont {A.}~\bibnamefont
  {Tsintzis}}, \bibinfo {author} {\bibfnamefont {M.}~\bibnamefont {Leijnse}},\
  and\ \bibinfo {author} {\bibfnamefont {J.}~\bibnamefont {Danon}},\ }\bibfield
   {title} {\bibinfo {title} {Probing {M}ajorana localization in minimal
  {K}itaev chains through a quantum dot},\ }\href
  {https://doi.org/10.1103/PhysRevResearch.5.043182} {\bibfield  {journal}
  {\bibinfo  {journal} {Phys. Rev. Res.}\ }\textbf {\bibinfo {volume} {5}},\
  \bibinfo {pages} {043182} (\bibinfo {year} {2023})}\BibitemShut {NoStop}%
\bibitem [{\citenamefont {Tsintzis}\ \emph {et~al.}(2024)\citenamefont
  {Tsintzis}, \citenamefont {Souto}, \citenamefont {Flensberg}, \citenamefont
  {Danon},\ and\ \citenamefont {Leijnse}}]{PRXQuantum.5.010323}%
  \BibitemOpen
  \bibfield  {author} {\bibinfo {author} {\bibfnamefont {A.}~\bibnamefont
  {Tsintzis}}, \bibinfo {author} {\bibfnamefont {R.~S.}\ \bibnamefont {Souto}},
  \bibinfo {author} {\bibfnamefont {K.}~\bibnamefont {Flensberg}}, \bibinfo
  {author} {\bibfnamefont {J.}~\bibnamefont {Danon}},\ and\ \bibinfo {author}
  {\bibfnamefont {M.}~\bibnamefont {Leijnse}},\ }\bibfield  {title} {\bibinfo
  {title} {Majorana qubits and non-{A}belian physics in quantum dot--based
  minimal {K}itaev chains},\ }\href
  {https://doi.org/10.1103/PRXQuantum.5.010323} {\bibfield  {journal} {\bibinfo
   {journal} {PRX Quantum}\ }\textbf {\bibinfo {volume} {5}},\ \bibinfo {pages}
  {010323} (\bibinfo {year} {2024})}\BibitemShut {NoStop}%
\bibitem [{\citenamefont {Liu}\ \emph {et~al.}(2024)\citenamefont {Liu},
  \citenamefont {Bozkurt}, \citenamefont {Zatelli}, \citenamefont {ten Haaf},
  \citenamefont {Dvir},\ and\ \citenamefont {Wimmer}}]{liu2024enhancing}%
  \BibitemOpen
  \bibfield  {author} {\bibinfo {author} {\bibfnamefont {C.-X.}\ \bibnamefont
  {Liu}}, \bibinfo {author} {\bibfnamefont {A.~M.}\ \bibnamefont {Bozkurt}},
  \bibinfo {author} {\bibfnamefont {F.}~\bibnamefont {Zatelli}}, \bibinfo
  {author} {\bibfnamefont {S.~L.}\ \bibnamefont {ten Haaf}}, \bibinfo {author}
  {\bibfnamefont {T.}~\bibnamefont {Dvir}},\ and\ \bibinfo {author}
  {\bibfnamefont {M.}~\bibnamefont {Wimmer}},\ }\bibfield  {title} {\bibinfo
  {title} {Enhancing the excitation gap of a quantum-dot-based {K}itaev
  chain},\ }\href {https://www.nature.com/articles/s42005-024-01715-5}
  {\bibfield  {journal} {\bibinfo  {journal} {Commun. Phys.}\ }\textbf
  {\bibinfo {volume} {7}},\ \bibinfo {pages} {235} (\bibinfo {year}
  {2024})}\BibitemShut {NoStop}%
\bibitem [{\citenamefont {Alvarado}\ \emph {et~al.}(2024)\citenamefont
  {Alvarado}, \citenamefont {Yeyati}, \citenamefont {Aguado},\ and\
  \citenamefont {Souto}}]{PhysRevB.110.245144}%
  \BibitemOpen
  \bibfield  {author} {\bibinfo {author} {\bibfnamefont {M.}~\bibnamefont
  {Alvarado}}, \bibinfo {author} {\bibfnamefont {A.~L.}\ \bibnamefont
  {Yeyati}}, \bibinfo {author} {\bibfnamefont {R.}~\bibnamefont {Aguado}},\
  and\ \bibinfo {author} {\bibfnamefont {R.~S.}\ \bibnamefont {Souto}},\
  }\bibfield  {title} {\bibinfo {title} {Interplay between {M}ajorana and
  {S}hiba states in a minimal {K}itaev chain coupled to a superconductor},\
  }\href {https://doi.org/10.1103/PhysRevB.110.245144} {\bibfield  {journal}
  {\bibinfo  {journal} {Phys. Rev. B}\ }\textbf {\bibinfo {volume} {110}},\
  \bibinfo {pages} {245144} (\bibinfo {year} {2024})}\BibitemShut {NoStop}%
\bibitem [{\citenamefont {Samuelson}\ \emph {et~al.}(2024)\citenamefont
  {Samuelson}, \citenamefont {Svensson},\ and\ \citenamefont
  {Leijnse}}]{PhysRevB.109.035415}%
  \BibitemOpen
  \bibfield  {author} {\bibinfo {author} {\bibfnamefont {W.}~\bibnamefont
  {Samuelson}}, \bibinfo {author} {\bibfnamefont {V.}~\bibnamefont
  {Svensson}},\ and\ \bibinfo {author} {\bibfnamefont {M.}~\bibnamefont
  {Leijnse}},\ }\bibfield  {title} {\bibinfo {title} {Minimal quantum dot based
  {K}itaev chain with only local superconducting proximity effect},\ }\href
  {https://doi.org/10.1103/PhysRevB.109.035415} {\bibfield  {journal} {\bibinfo
   {journal} {Phys. Rev. B}\ }\textbf {\bibinfo {volume} {109}},\ \bibinfo
  {pages} {035415} (\bibinfo {year} {2024})}\BibitemShut {NoStop}%
\bibitem [{\citenamefont {Awoga}\ and\ \citenamefont
  {Cayao}(2024)}]{PhysRevB.110.165404}%
  \BibitemOpen
  \bibfield  {author} {\bibinfo {author} {\bibfnamefont {O.~A.}\ \bibnamefont
  {Awoga}}\ and\ \bibinfo {author} {\bibfnamefont {J.}~\bibnamefont {Cayao}},\
  }\bibfield  {title} {\bibinfo {title} {Identifying trivial and {M}ajorana
  zero-energy modes using the {M}ajorana polarization},\ }\href
  {https://doi.org/10.1103/PhysRevB.110.165404} {\bibfield  {journal} {\bibinfo
   {journal} {Phys. Rev. B}\ }\textbf {\bibinfo {volume} {110}},\ \bibinfo
  {pages} {165404} (\bibinfo {year} {2024})}\BibitemShut {NoStop}%
\bibitem [{\citenamefont {Luethi}\ \emph {et~al.}(2024)\citenamefont {Luethi},
  \citenamefont {Legg}, \citenamefont {Loss},\ and\ \citenamefont
  {Klinovaja}}]{PhysRevB.110.245412}%
  \BibitemOpen
  \bibfield  {author} {\bibinfo {author} {\bibfnamefont {M.}~\bibnamefont
  {Luethi}}, \bibinfo {author} {\bibfnamefont {H.~F.}\ \bibnamefont {Legg}},
  \bibinfo {author} {\bibfnamefont {D.}~\bibnamefont {Loss}},\ and\ \bibinfo
  {author} {\bibfnamefont {J.}~\bibnamefont {Klinovaja}},\ }\bibfield  {title}
  {\bibinfo {title} {From perfect to imperfect poor man's {M}ajoranas in
  minimal {K}itaev chains},\ }\href
  {https://doi.org/10.1103/PhysRevB.110.245412} {\bibfield  {journal} {\bibinfo
   {journal} {Phys. Rev. B}\ }\textbf {\bibinfo {volume} {110}},\ \bibinfo
  {pages} {245412} (\bibinfo {year} {2024})}\BibitemShut {NoStop}%
\bibitem [{\citenamefont {Nitsch}\ \emph {et~al.}(2024)\citenamefont {Nitsch},
  \citenamefont {Maffi}, \citenamefont {Baran}, \citenamefont {Souto},
  \citenamefont {Paaske}, \citenamefont {Leijnse},\ and\ \citenamefont
  {Burrello}}]{nitsch2024tetron}%
  \BibitemOpen
  \bibfield  {author} {\bibinfo {author} {\bibfnamefont {M.}~\bibnamefont
  {Nitsch}}, \bibinfo {author} {\bibfnamefont {L.}~\bibnamefont {Maffi}},
  \bibinfo {author} {\bibfnamefont {V.~V.}\ \bibnamefont {Baran}}, \bibinfo
  {author} {\bibfnamefont {R.~S.}\ \bibnamefont {Souto}}, \bibinfo {author}
  {\bibfnamefont {J.}~\bibnamefont {Paaske}}, \bibinfo {author} {\bibfnamefont
  {M.}~\bibnamefont {Leijnse}},\ and\ \bibinfo {author} {\bibfnamefont
  {M.}~\bibnamefont {Burrello}},\ }\bibfield  {title} {\bibinfo {title} {The
  poor man's majorana tetron},\ }\href {https://arxiv.org/abs/2411.11981}
  {\bibfield  {journal} {\bibinfo  {journal} {arXiv:2411.11981}\ } (\bibinfo
  {year} {2024})}\BibitemShut {NoStop}%
\bibitem [{\citenamefont {Kotetes}\ \emph {et~al.}(2024)\citenamefont
  {Kotetes}, \citenamefont {Roig},\ and\ \citenamefont
  {Andersen}}]{kotetes2024nonRecifourpi}%
  \BibitemOpen
  \bibfield  {author} {\bibinfo {author} {\bibfnamefont {P.}~\bibnamefont
  {Kotetes}}, \bibinfo {author} {\bibfnamefont {M.}~\bibnamefont {Roig}},\ and\
  \bibinfo {author} {\bibfnamefont {B.~M.}\ \bibnamefont {Andersen}},\
  }\bibfield  {title} {\bibinfo {title} {Nonreciprocal equilibrium
  4$\pi$-periodic {J}osephson effect from poor man's {M}ajorana zero modes},\
  }\href {https://arxiv.org/abs/2409.13027} {\bibfield  {journal} {\bibinfo
  {journal} {arXiv:2409.13027}\ } (\bibinfo {year} {2024})}\BibitemShut
  {NoStop}%
\bibitem [{\citenamefont {Kumar}\ \emph {et~al.}(2026)\citenamefont {Kumar},
  \citenamefont {Cayao},\ and\ \citenamefont {Awoga}}]{kumar2026Awoga}%
  \BibitemOpen
  \bibfield  {author} {\bibinfo {author} {\bibfnamefont {A.~S.}\ \bibnamefont
  {Kumar}}, \bibinfo {author} {\bibfnamefont {J.}~\bibnamefont {Cayao}},\ and\
  \bibinfo {author} {\bibfnamefont {O.~A.}\ \bibnamefont {Awoga}},\ }\bibfield
  {title} {\bibinfo {title} {Nonlocal {M}ajorana polarization in
  non-{H}ermitian topological superconductors},\ }\href
  {https://arxiv.org/abs/2607.25424} {\bibfield  {journal} {\bibinfo  {journal}
  {arXiv:2607.25424}\ } (\bibinfo {year} {2026})}\BibitemShut {NoStop}%
\bibitem [{\citenamefont {Luethi}\ \emph
  {et~al.}(2025{\natexlab{a}})\citenamefont {Luethi}, \citenamefont {Legg},
  \citenamefont {Loss},\ and\ \citenamefont {Klinovaja}}]{PhysRevB.111.115419}%
  \BibitemOpen
  \bibfield  {author} {\bibinfo {author} {\bibfnamefont {M.}~\bibnamefont
  {Luethi}}, \bibinfo {author} {\bibfnamefont {H.~F.}\ \bibnamefont {Legg}},
  \bibinfo {author} {\bibfnamefont {D.}~\bibnamefont {Loss}},\ and\ \bibinfo
  {author} {\bibfnamefont {J.}~\bibnamefont {Klinovaja}},\ }\bibfield  {title}
  {\bibinfo {title} {Fate of poor man's {M}ajoranas in the long {K}itaev chain
  limit},\ }\href {https://doi.org/10.1103/PhysRevB.111.115419} {\bibfield
  {journal} {\bibinfo  {journal} {Phys. Rev. B}\ }\textbf {\bibinfo {volume}
  {111}},\ \bibinfo {pages} {115419} (\bibinfo {year}
  {2025}{\natexlab{a}})}\BibitemShut {NoStop}%
\bibitem [{\citenamefont {Vimal}\ and\ \citenamefont
  {Cayao}(2025)}]{vimal2025EntMKC}%
  \BibitemOpen
  \bibfield  {author} {\bibinfo {author} {\bibfnamefont {V.~K.}\ \bibnamefont
  {Vimal}}\ and\ \bibinfo {author} {\bibfnamefont {J.}~\bibnamefont {Cayao}},\
  }\bibfield  {title} {\bibinfo {title} {Entanglement dynamics in minimal
  {K}itaev chains},\ }\href {https://arxiv.org/abs/2507.17586} {\bibfield
  {journal} {\bibinfo  {journal} {arXiv: 2507.17586}\ } (\bibinfo {year}
  {2025})}\BibitemShut {NoStop}%
\bibitem [{\citenamefont {Seoane~Souto}\ and\ \citenamefont
  {Aguado}(2024)}]{Seoane_Souto_2024}%
  \BibitemOpen
  \bibfield  {author} {\bibinfo {author} {\bibfnamefont {R.}~\bibnamefont
  {Seoane~Souto}}\ and\ \bibinfo {author} {\bibfnamefont {R.}~\bibnamefont
  {Aguado}},\ }\bibinfo {title} {Subgap states in semiconductor-superconductor
  devices for quantum technologies: {A}ndreev qubits and minimal {M}ajorana
  chains},\ in\ \href {https://doi.org/10.1007/978-3-031-55657-9_3} {\emph
  {\bibinfo {booktitle} {New Trends and Platforms for Quantum Technologies}}}\
  (\bibinfo  {publisher} {Springer Nature Switzerland},\ \bibinfo {year}
  {2024})\ p.\ \bibinfo {pages} {133–223}\BibitemShut {NoStop}%
\bibitem [{\citenamefont {Pita-Vidal}\ \emph {et~al.}(2025)\citenamefont
  {Pita-Vidal}, \citenamefont {Souto}, \citenamefont {Goswami}, \citenamefont
  {Andersen}, \citenamefont {Katsaros}, \citenamefont {Shabani},\ and\
  \citenamefont {Aguado}}]{pitavidal2025qubit}%
  \BibitemOpen
  \bibfield  {author} {\bibinfo {author} {\bibfnamefont {M.}~\bibnamefont
  {Pita-Vidal}}, \bibinfo {author} {\bibfnamefont {R.~S.}\ \bibnamefont
  {Souto}}, \bibinfo {author} {\bibfnamefont {S.}~\bibnamefont {Goswami}},
  \bibinfo {author} {\bibfnamefont {C.~K.}\ \bibnamefont {Andersen}}, \bibinfo
  {author} {\bibfnamefont {G.}~\bibnamefont {Katsaros}}, \bibinfo {author}
  {\bibfnamefont {J.}~\bibnamefont {Shabani}},\ and\ \bibinfo {author}
  {\bibfnamefont {R.}~\bibnamefont {Aguado}},\ }\bibfield  {title} {\bibinfo
  {title} {Novel qubits in hybrid semiconductor-superconductor
  nanostructures},\ }\href {https://arxiv.org/abs/2512.23336} {\bibfield
  {journal} {\bibinfo  {journal} {arXiv:2512.23336}\ } (\bibinfo {year}
  {2025})}\BibitemShut {NoStop}%
\bibitem [{\citenamefont {Alvarado}\ \emph {et~al.}(2026)\citenamefont
  {Alvarado}, \citenamefont {Souto}, \citenamefont {Calder\'on},\ and\
  \citenamefont {Aguado}}]{w8f7-6cyy}%
  \BibitemOpen
  \bibfield  {author} {\bibinfo {author} {\bibfnamefont {M.}~\bibnamefont
  {Alvarado}}, \bibinfo {author} {\bibfnamefont {R.~S.}\ \bibnamefont {Souto}},
  \bibinfo {author} {\bibfnamefont {M.~J.}\ \bibnamefont {Calder\'on}},\ and\
  \bibinfo {author} {\bibfnamefont {R.}~\bibnamefont {Aguado}},\ }\bibfield
  {title} {\bibinfo {title} {Optimal majoranas in mesoscopic {K}itaev chains},\
  }\href {https://doi.org/10.1103/w8f7-6cyy} {\bibfield  {journal} {\bibinfo
  {journal} {Phys. Rev. B}\ }\textbf {\bibinfo {volume} {114}},\ \bibinfo
  {pages} {034512} (\bibinfo {year} {2026})}\BibitemShut {NoStop}%
\bibitem [{\citenamefont {Alvarado}\ \emph {et~al.}(2025)\citenamefont
  {Alvarado}, \citenamefont {Yeyati}, \citenamefont {Aguado},\ and\
  \citenamefont {Souto}}]{1jqt-3ht5}%
  \BibitemOpen
  \bibfield  {author} {\bibinfo {author} {\bibfnamefont {M.}~\bibnamefont
  {Alvarado}}, \bibinfo {author} {\bibfnamefont {A.~L.}\ \bibnamefont
  {Yeyati}}, \bibinfo {author} {\bibfnamefont {R.}~\bibnamefont {Aguado}},\
  and\ \bibinfo {author} {\bibfnamefont {R.~S.}\ \bibnamefont {Souto}},\
  }\bibfield  {title} {\bibinfo {title} {Characterizing local majorana
  properties using {A}ndreev states},\ }\href
  {https://doi.org/10.1103/1jqt-3ht5} {\bibfield  {journal} {\bibinfo
  {journal} {Phys. Rev. B}\ }\textbf {\bibinfo {volume} {112}},\ \bibinfo
  {pages} {L241403} (\bibinfo {year} {2025})}\BibitemShut {NoStop}%
\bibitem [{\citenamefont {Luethi}\ \emph
  {et~al.}(2025{\natexlab{b}})\citenamefont {Luethi}, \citenamefont {Legg},
  \citenamefont {Loss},\ and\ \citenamefont {Klinovaja}}]{bg4z-vx9m}%
  \BibitemOpen
  \bibfield  {author} {\bibinfo {author} {\bibfnamefont {M.}~\bibnamefont
  {Luethi}}, \bibinfo {author} {\bibfnamefont {H.~F.}\ \bibnamefont {Legg}},
  \bibinfo {author} {\bibfnamefont {D.}~\bibnamefont {Loss}},\ and\ \bibinfo
  {author} {\bibfnamefont {J.}~\bibnamefont {Klinovaja}},\ }\bibfield  {title}
  {\bibinfo {title} {Properties and prevalence of false poor man's {M}ajoranas
  in two- and three-site artificial {K}itaev chains},\ }\href
  {https://doi.org/10.1103/bg4z-vx9m} {\bibfield  {journal} {\bibinfo
  {journal} {Phys. Rev. B}\ }\textbf {\bibinfo {volume} {112}},\ \bibinfo
  {pages} {205426} (\bibinfo {year} {2025}{\natexlab{b}})}\BibitemShut
  {NoStop}%
\bibitem [{\citenamefont {Cayao}\ and\ \citenamefont
  {Sato}(2026{\natexlab{a}})}]{hf7s-f7tj}%
  \BibitemOpen
  \bibfield  {author} {\bibinfo {author} {\bibfnamefont {J.}~\bibnamefont
  {Cayao}}\ and\ \bibinfo {author} {\bibfnamefont {M.}~\bibnamefont {Sato}},\
  }\bibfield  {title} {\bibinfo {title} {Nonlocal {J}osephson diode effect in
  minimal {K}itaev chains},\ }\href {https://doi.org/10.1103/hf7s-f7tj}
  {\bibfield  {journal} {\bibinfo  {journal} {Phys. Rev. Res.}\ }\textbf
  {\bibinfo {volume} {8}},\ \bibinfo {pages} {013326} (\bibinfo {year}
  {2026}{\natexlab{a}})}\BibitemShut {NoStop}%
\bibitem [{\citenamefont {Cayao}\ and\ \citenamefont
  {Sato}()}]{cayaosatosymmetryMKC}%
  \BibitemOpen
  \bibfield  {author} {\bibinfo {author} {\bibfnamefont {J.}~\bibnamefont
  {Cayao}}\ and\ \bibinfo {author} {\bibfnamefont {M.}~\bibnamefont {Sato}},\
  }\href@noop {} {\bibinfo {title} {To be published elsewhere}}\BibitemShut
  {NoStop}%
\bibitem [{\citenamefont {Datta}(1997)}]{datta1997electronic}%
  \BibitemOpen
  \bibfield  {author} {\bibinfo {author} {\bibfnamefont {S.}~\bibnamefont
  {Datta}},\ }\href@noop {} {\emph {\bibinfo {title} {Electronic transport in
  mesoscopic systems}}}\ (\bibinfo  {publisher} {Cambridge university press},\
  \bibinfo {year} {1997})\BibitemShut {NoStop}%
\bibitem [{\citenamefont {Dembowski}\ \emph {et~al.}(2001)\citenamefont
  {Dembowski}, \citenamefont {Gr\"af}, \citenamefont {Harney}, \citenamefont
  {Heine}, \citenamefont {Heiss}, \citenamefont {Rehfeld},\ and\ \citenamefont
  {Richter}}]{PhysRevLett.86.787}%
  \BibitemOpen
  \bibfield  {author} {\bibinfo {author} {\bibfnamefont {C.}~\bibnamefont
  {Dembowski}}, \bibinfo {author} {\bibfnamefont {H.-D.}\ \bibnamefont
  {Gr\"af}}, \bibinfo {author} {\bibfnamefont {H.~L.}\ \bibnamefont {Harney}},
  \bibinfo {author} {\bibfnamefont {A.}~\bibnamefont {Heine}}, \bibinfo
  {author} {\bibfnamefont {W.~D.}\ \bibnamefont {Heiss}}, \bibinfo {author}
  {\bibfnamefont {H.}~\bibnamefont {Rehfeld}},\ and\ \bibinfo {author}
  {\bibfnamefont {A.}~\bibnamefont {Richter}},\ }\bibfield  {title} {\bibinfo
  {title} {Experimental observation of the topological structure of exceptional
  points},\ }\href {https://doi.org/10.1103/PhysRevLett.86.787} {\bibfield
  {journal} {\bibinfo  {journal} {Phys. Rev. Lett.}\ }\textbf {\bibinfo
  {volume} {86}},\ \bibinfo {pages} {787} (\bibinfo {year} {2001})}\BibitemShut
  {NoStop}%
\bibitem [{\citenamefont {Kawabata}\ \emph
  {et~al.}(2019{\natexlab{a}})\citenamefont {Kawabata}, \citenamefont
  {Shiozaki}, \citenamefont {Ueda},\ and\ \citenamefont
  {Sato}}]{PhysRevX.9.041015}%
  \BibitemOpen
  \bibfield  {author} {\bibinfo {author} {\bibfnamefont {K.}~\bibnamefont
  {Kawabata}}, \bibinfo {author} {\bibfnamefont {K.}~\bibnamefont {Shiozaki}},
  \bibinfo {author} {\bibfnamefont {M.}~\bibnamefont {Ueda}},\ and\ \bibinfo
  {author} {\bibfnamefont {M.}~\bibnamefont {Sato}},\ }\bibfield  {title}
  {\bibinfo {title} {Symmetry and topology in non-{H}ermitian physics},\ }\href
  {https://doi.org/10.1103/PhysRevX.9.041015} {\bibfield  {journal} {\bibinfo
  {journal} {Phys. Rev. X}\ }\textbf {\bibinfo {volume} {9}},\ \bibinfo {pages}
  {041015} (\bibinfo {year} {2019}{\natexlab{a}})}\BibitemShut {NoStop}%
\bibitem [{\citenamefont {Gong}\ \emph {et~al.}(2018)\citenamefont {Gong},
  \citenamefont {Ashida}, \citenamefont {Kawabata}, \citenamefont {Takasan},
  \citenamefont {Higashikawa},\ and\ \citenamefont {Ueda}}]{PhysRevX.8.031079}%
  \BibitemOpen
  \bibfield  {author} {\bibinfo {author} {\bibfnamefont {Z.}~\bibnamefont
  {Gong}}, \bibinfo {author} {\bibfnamefont {Y.}~\bibnamefont {Ashida}},
  \bibinfo {author} {\bibfnamefont {K.}~\bibnamefont {Kawabata}}, \bibinfo
  {author} {\bibfnamefont {K.}~\bibnamefont {Takasan}}, \bibinfo {author}
  {\bibfnamefont {S.}~\bibnamefont {Higashikawa}},\ and\ \bibinfo {author}
  {\bibfnamefont {M.}~\bibnamefont {Ueda}},\ }\bibfield  {title} {\bibinfo
  {title} {Topological phases of non-{H}ermitian systems},\ }\href
  {https://doi.org/10.1103/PhysRevX.8.031079} {\bibfield  {journal} {\bibinfo
  {journal} {Phys. Rev. X}\ }\textbf {\bibinfo {volume} {8}},\ \bibinfo {pages}
  {031079} (\bibinfo {year} {2018})}\BibitemShut {NoStop}%
\bibitem [{\citenamefont {Bessho}\ \emph {et~al.}(2019)\citenamefont {Bessho},
  \citenamefont {Kawabata},\ and\ \citenamefont
  {Sato}}]{doi:10.7566/JPSCP.30.011098}%
  \BibitemOpen
  \bibfield  {author} {\bibinfo {author} {\bibfnamefont {T.}~\bibnamefont
  {Bessho}}, \bibinfo {author} {\bibfnamefont {K.}~\bibnamefont {Kawabata}},\
  and\ \bibinfo {author} {\bibfnamefont {M.}~\bibnamefont {Sato}},\ }\bibinfo
  {title} {Topological classificaton of non-{H}ermitian gapless phases:
  Exceptional points and bulk fermi arcs},\ in\ \href
  {https://doi.org/10.7566/JPSCP.30.011098} {\emph {\bibinfo {booktitle} {Proc.
  Int. Conf. on Strongly Correlated Electron Systems (SCES2019)}}}\ (\bibinfo
  {publisher} {Physical Society of Japan},\ \bibinfo {year} {2019})\
  Chap.~\bibinfo {chapter} {30}, p.\ \bibinfo {pages} {011098}\BibitemShut
  {NoStop}%
\bibitem [{\citenamefont {Okuma}\ and\ \citenamefont {Sato}(2023)}]{OS23}%
  \BibitemOpen
  \bibfield  {author} {\bibinfo {author} {\bibfnamefont {N.}~\bibnamefont
  {Okuma}}\ and\ \bibinfo {author} {\bibfnamefont {M.}~\bibnamefont {Sato}},\
  }\bibfield  {title} {\bibinfo {title} {Non-{H}ermitian topological phenomena:
  a review},\ }\href {https://doi.org/10.1146/annurev-conmatphys-040521-033133}
  {\bibfield  {journal} {\bibinfo  {journal} {Annu. Rev. Condens. Matter
  Phys.}\ ,\ \bibinfo {pages} {83}} (\bibinfo {year} {2023})}\BibitemShut
  {NoStop}%
\bibitem [{\citenamefont {Bergholtz}\ \emph {et~al.}(2021)\citenamefont
  {Bergholtz}, \citenamefont {Budich},\ and\ \citenamefont
  {Kunst}}]{RevModPhys.93.015005}%
  \BibitemOpen
  \bibfield  {author} {\bibinfo {author} {\bibfnamefont {E.~J.}\ \bibnamefont
  {Bergholtz}}, \bibinfo {author} {\bibfnamefont {J.~C.}\ \bibnamefont
  {Budich}},\ and\ \bibinfo {author} {\bibfnamefont {F.~K.}\ \bibnamefont
  {Kunst}},\ }\bibfield  {title} {\bibinfo {title} {Exceptional topology of
  non-{H}ermitian systems},\ }\href
  {https://doi.org/10.1103/RevModPhys.93.015005} {\bibfield  {journal}
  {\bibinfo  {journal} {Rev. Mod. Phys.}\ }\textbf {\bibinfo {volume} {93}},\
  \bibinfo {pages} {015005} (\bibinfo {year} {2021})}\BibitemShut {NoStop}%
\bibitem [{\citenamefont {El-Ganainy}\ \emph {et~al.}(2018)\citenamefont
  {El-Ganainy}, \citenamefont {Makris}, \citenamefont {Khajavikhan},
  \citenamefont {Musslimani}, \citenamefont {Rotter},\ and\ \citenamefont
  {Christodoulides}}]{el2018non}%
  \BibitemOpen
  \bibfield  {author} {\bibinfo {author} {\bibfnamefont {R.}~\bibnamefont
  {El-Ganainy}}, \bibinfo {author} {\bibfnamefont {K.~G.}\ \bibnamefont
  {Makris}}, \bibinfo {author} {\bibfnamefont {M.}~\bibnamefont {Khajavikhan}},
  \bibinfo {author} {\bibfnamefont {Z.~H.}\ \bibnamefont {Musslimani}},
  \bibinfo {author} {\bibfnamefont {S.}~\bibnamefont {Rotter}},\ and\ \bibinfo
  {author} {\bibfnamefont {D.~N.}\ \bibnamefont {Christodoulides}},\ }\bibfield
   {title} {\bibinfo {title} {Non-{H}ermitian physics and {PT} symmetry},\
  }\href {https://www.nature.com/articles/nphys4323} {\bibfield  {journal}
  {\bibinfo  {journal} {Nat. Phys.}\ }\textbf {\bibinfo {volume} {14}},\
  \bibinfo {pages} {11} (\bibinfo {year} {2018})}\BibitemShut {NoStop}%
\bibitem [{\citenamefont {{\"O}zdemir}\ \emph {et~al.}(2019)\citenamefont
  {{\"O}zdemir}, \citenamefont {Rotter}, \citenamefont {Nori},\ and\
  \citenamefont {Yang}}]{ozdemir2019parity}%
  \BibitemOpen
  \bibfield  {author} {\bibinfo {author} {\bibfnamefont {{\c{S}}.~K.}\
  \bibnamefont {{\"O}zdemir}}, \bibinfo {author} {\bibfnamefont
  {S.}~\bibnamefont {Rotter}}, \bibinfo {author} {\bibfnamefont
  {F.}~\bibnamefont {Nori}},\ and\ \bibinfo {author} {\bibfnamefont
  {L.}~\bibnamefont {Yang}},\ }\bibfield  {title} {\bibinfo {title}
  {Parity--time symmetry and exceptional points in photonics},\ }\href
  {https://www.nature.com/articles/s41563-019-0304-9} {\bibfield  {journal}
  {\bibinfo  {journal} {Nat. Mater.}\ }\textbf {\bibinfo {volume} {18}},\
  \bibinfo {pages} {783} (\bibinfo {year} {2019})}\BibitemShut {NoStop}%
\bibitem [{\citenamefont {Ashida}\ \emph {et~al.}(2020)\citenamefont {Ashida},
  \citenamefont {Gong},\ and\ \citenamefont
  {Ueda}}]{doi:10.1080/00018732.2021.1876991}%
  \BibitemOpen
  \bibfield  {author} {\bibinfo {author} {\bibfnamefont {Y.}~\bibnamefont
  {Ashida}}, \bibinfo {author} {\bibfnamefont {Z.}~\bibnamefont {Gong}},\ and\
  \bibinfo {author} {\bibfnamefont {M.}~\bibnamefont {Ueda}},\ }\bibfield
  {title} {\bibinfo {title} {Non-{H}ermitian physics},\ }\href
  {https://doi.org/10.1080/00018732.2021.1876991} {\bibfield  {journal}
  {\bibinfo  {journal} {Adv. Phys.}\ }\textbf {\bibinfo {volume} {69}},\
  \bibinfo {pages} {249} (\bibinfo {year} {2020})}\BibitemShut {NoStop}%
\bibitem [{\citenamefont {San-Jos\'{e}}\ \emph {et~al.}(2016)\citenamefont
  {San-Jos\'{e}}, \citenamefont {Cayao}, \citenamefont {Prada},\ and\
  \citenamefont {Aguado}}]{JorgeEPs}%
  \BibitemOpen
  \bibfield  {author} {\bibinfo {author} {\bibfnamefont {P.}~\bibnamefont
  {San-Jos\'{e}}}, \bibinfo {author} {\bibfnamefont {J.}~\bibnamefont {Cayao}},
  \bibinfo {author} {\bibfnamefont {E.}~\bibnamefont {Prada}},\ and\ \bibinfo
  {author} {\bibfnamefont {R.}~\bibnamefont {Aguado}},\ }\bibfield  {title}
  {\bibinfo {title} {Majorana bound states from exceptional points in
  non-topological superconductors},\ }\href
  {http://dx.doi.org/10.1038/srep21427} {\bibfield  {journal} {\bibinfo
  {journal} {Sci. Rep.}\ }\textbf {\bibinfo {volume} {6}},\ \bibinfo {pages}
  {21427} (\bibinfo {year} {2016})}\BibitemShut {NoStop}%
\bibitem [{\citenamefont {Avila}\ \emph {et~al.}(2019)\citenamefont {Avila},
  \citenamefont {Pe{\~n}aranda}, \citenamefont {Prada}, \citenamefont
  {San-Jose},\ and\ \citenamefont {Aguado}}]{avila2019non}%
  \BibitemOpen
  \bibfield  {author} {\bibinfo {author} {\bibfnamefont {J.}~\bibnamefont
  {Avila}}, \bibinfo {author} {\bibfnamefont {F.}~\bibnamefont
  {Pe{\~n}aranda}}, \bibinfo {author} {\bibfnamefont {E.}~\bibnamefont
  {Prada}}, \bibinfo {author} {\bibfnamefont {P.}~\bibnamefont {San-Jose}},\
  and\ \bibinfo {author} {\bibfnamefont {R.}~\bibnamefont {Aguado}},\
  }\bibfield  {title} {\bibinfo {title} {Non-{H}ermitian topology as a unifying
  framework for the {A}ndreev versus {M}ajorana states controversy},\ }\href
  {https://doi.org/10.1038/s42005-019-0231-8} {\bibfield  {journal} {\bibinfo
  {journal} {Commun. Phys.}\ }\textbf {\bibinfo {volume} {2}},\ \bibinfo
  {pages} {133} (\bibinfo {year} {2019})}\BibitemShut {NoStop}%
\bibitem [{\citenamefont {Okuma}\ and\ \citenamefont
  {Sato}(2019)}]{PhysRevLett.123.097701}%
  \BibitemOpen
  \bibfield  {author} {\bibinfo {author} {\bibfnamefont {N.}~\bibnamefont
  {Okuma}}\ and\ \bibinfo {author} {\bibfnamefont {M.}~\bibnamefont {Sato}},\
  }\bibfield  {title} {\bibinfo {title} {Topological phase transition driven by
  infinitesimal instability: {M}ajorana fermions in non-{H}ermitian
  spintronics},\ }\href {https://doi.org/10.1103/PhysRevLett.123.097701}
  {\bibfield  {journal} {\bibinfo  {journal} {Phys. Rev. Lett.}\ }\textbf
  {\bibinfo {volume} {123}},\ \bibinfo {pages} {097701} (\bibinfo {year}
  {2019})}\BibitemShut {NoStop}%
\bibitem [{\citenamefont {Okugawa}\ and\ \citenamefont
  {Yokoyama}(2019)}]{PhysRevB.99.041202}%
  \BibitemOpen
  \bibfield  {author} {\bibinfo {author} {\bibfnamefont {R.}~\bibnamefont
  {Okugawa}}\ and\ \bibinfo {author} {\bibfnamefont {T.}~\bibnamefont
  {Yokoyama}},\ }\bibfield  {title} {\bibinfo {title} {Topological exceptional
  surfaces in non-{H}ermitian systems with parity-time and parity-particle-hole
  symmetries},\ }\href {https://doi.org/10.1103/PhysRevB.99.041202} {\bibfield
  {journal} {\bibinfo  {journal} {Phys. Rev. B}\ }\textbf {\bibinfo {volume}
  {99}},\ \bibinfo {pages} {041202} (\bibinfo {year} {2019})}\BibitemShut
  {NoStop}%
\bibitem [{\citenamefont {Zyuzin}\ and\ \citenamefont
  {Simon}(2019)}]{PhysRevB.99.165145}%
  \BibitemOpen
  \bibfield  {author} {\bibinfo {author} {\bibfnamefont {A.~A.}\ \bibnamefont
  {Zyuzin}}\ and\ \bibinfo {author} {\bibfnamefont {P.}~\bibnamefont {Simon}},\
  }\bibfield  {title} {\bibinfo {title} {Disorder-induced exceptional points
  and nodal lines in dirac superconductors},\ }\href
  {https://doi.org/10.1103/PhysRevB.99.165145} {\bibfield  {journal} {\bibinfo
  {journal} {Phys. Rev. B}\ }\textbf {\bibinfo {volume} {99}},\ \bibinfo
  {pages} {165145} (\bibinfo {year} {2019})}\BibitemShut {NoStop}%
\bibitem [{\citenamefont {Cayao}\ and\ \citenamefont
  {Black-Schaffer}(2022)}]{PhysRevB.105.094502}%
  \BibitemOpen
  \bibfield  {author} {\bibinfo {author} {\bibfnamefont {J.}~\bibnamefont
  {Cayao}}\ and\ \bibinfo {author} {\bibfnamefont {A.~M.}\ \bibnamefont
  {Black-Schaffer}},\ }\bibfield  {title} {\bibinfo {title} {Exceptional
  odd-frequency pairing in non-{H}ermitian superconducting systems},\ }\href
  {https://doi.org/10.1103/PhysRevB.105.094502} {\bibfield  {journal} {\bibinfo
   {journal} {Phys. Rev. B}\ }\textbf {\bibinfo {volume} {105}},\ \bibinfo
  {pages} {094502} (\bibinfo {year} {2022})}\BibitemShut {NoStop}%
\bibitem [{\citenamefont {Cayao}\ and\ \citenamefont
  {Black-Schaffer}(2023)}]{PhysRevB.107.104515}%
  \BibitemOpen
  \bibfield  {author} {\bibinfo {author} {\bibfnamefont {J.}~\bibnamefont
  {Cayao}}\ and\ \bibinfo {author} {\bibfnamefont {A.~M.}\ \bibnamefont
  {Black-Schaffer}},\ }\bibfield  {title} {\bibinfo {title} {Bulk {B}ogoliubov
  {F}ermi arcs in non-{H}ermitian superconducting systems},\ }\href
  {https://doi.org/10.1103/PhysRevB.107.104515} {\bibfield  {journal} {\bibinfo
   {journal} {Phys. Rev. B}\ }\textbf {\bibinfo {volume} {107}},\ \bibinfo
  {pages} {104515} (\bibinfo {year} {2023})}\BibitemShut {NoStop}%
\bibitem [{\citenamefont {Cayao}(2024{\natexlab{b}})}]{PhysRevB.110.085414}%
  \BibitemOpen
  \bibfield  {author} {\bibinfo {author} {\bibfnamefont {J.}~\bibnamefont
  {Cayao}},\ }\bibfield  {title} {\bibinfo {title} {Non-{H}ermitian zero-energy
  pinning of {A}ndreev and {M}ajorana bound states in
  superconductor-semiconductor systems},\ }\href
  {https://doi.org/10.1103/PhysRevB.110.085414} {\bibfield  {journal} {\bibinfo
   {journal} {Phys. Rev. B}\ }\textbf {\bibinfo {volume} {110}},\ \bibinfo
  {pages} {085414} (\bibinfo {year} {2024}{\natexlab{b}})}\BibitemShut
  {NoStop}%
\bibitem [{\citenamefont {Cayao}\ and\ \citenamefont
  {Sato}(2024{\natexlab{a}})}]{cayao2023nonhermitian}%
  \BibitemOpen
  \bibfield  {author} {\bibinfo {author} {\bibfnamefont {J.}~\bibnamefont
  {Cayao}}\ and\ \bibinfo {author} {\bibfnamefont {M.}~\bibnamefont {Sato}},\
  }\bibfield  {title} {\bibinfo {title} {Non-{H}ermitian phase-biased
  {J}osephson junctions},\ }\href
  {https://doi.org/10.1103/PhysRevB.110.L201403} {\bibfield  {journal}
  {\bibinfo  {journal} {Phys. Rev. B}\ }\textbf {\bibinfo {volume} {110}},\
  \bibinfo {pages} {L201403} (\bibinfo {year}
  {2024}{\natexlab{a}})}\BibitemShut {NoStop}%
\bibitem [{\citenamefont {Li}\ \emph {et~al.}(2024)\citenamefont {Li},
  \citenamefont {Sun},\ and\ \citenamefont {Trauzettel}}]{li2023anomalous}%
  \BibitemOpen
  \bibfield  {author} {\bibinfo {author} {\bibfnamefont {C.-A.}\ \bibnamefont
  {Li}}, \bibinfo {author} {\bibfnamefont {H.-P.}\ \bibnamefont {Sun}},\ and\
  \bibinfo {author} {\bibfnamefont {B.}~\bibnamefont {Trauzettel}},\ }\bibfield
   {title} {\bibinfo {title} {Anomalous {A}ndreev spectrum and transport in
  non-{H}ermitian {J}osephson junctions},\ }\href
  {https://doi.org/10.1103/PhysRevB.109.214514} {\bibfield  {journal} {\bibinfo
   {journal} {Phys. Rev. B}\ }\textbf {\bibinfo {volume} {109}},\ \bibinfo
  {pages} {214514} (\bibinfo {year} {2024})}\BibitemShut {NoStop}%
\bibitem [{\citenamefont {Arouca}\ \emph {et~al.}(2023)\citenamefont {Arouca},
  \citenamefont {Cayao},\ and\ \citenamefont
  {Black-Schaffer}}]{PhysRevB.108.L060506}%
  \BibitemOpen
  \bibfield  {author} {\bibinfo {author} {\bibfnamefont {R.}~\bibnamefont
  {Arouca}}, \bibinfo {author} {\bibfnamefont {J.}~\bibnamefont {Cayao}},\ and\
  \bibinfo {author} {\bibfnamefont {A.~M.}\ \bibnamefont {Black-Schaffer}},\
  }\bibfield  {title} {\bibinfo {title} {Topological superconductivity enhanced
  by exceptional points},\ }\href
  {https://doi.org/10.1103/PhysRevB.108.L060506} {\bibfield  {journal}
  {\bibinfo  {journal} {Phys. Rev. B}\ }\textbf {\bibinfo {volume} {108}},\
  \bibinfo {pages} {L060506} (\bibinfo {year} {2023})}\BibitemShut {NoStop}%
\bibitem [{\citenamefont {Shen}\ \emph {et~al.}(2024)\citenamefont {Shen},
  \citenamefont {Lu}, \citenamefont {Lado},\ and\ \citenamefont
  {Trif}}]{shen2024nonhermitian}%
  \BibitemOpen
  \bibfield  {author} {\bibinfo {author} {\bibfnamefont {P.-X.}\ \bibnamefont
  {Shen}}, \bibinfo {author} {\bibfnamefont {Z.}~\bibnamefont {Lu}}, \bibinfo
  {author} {\bibfnamefont {J.~L.}\ \bibnamefont {Lado}},\ and\ \bibinfo
  {author} {\bibfnamefont {M.}~\bibnamefont {Trif}},\ }\bibfield  {title}
  {\bibinfo {title} {Non-{H}ermitian fermi-dirac distribution in persistent
  current transport},\ }\href {https://doi.org/10.1103/PhysRevLett.133.086301}
  {\bibfield  {journal} {\bibinfo  {journal} {Phys. Rev. Lett.}\ }\textbf
  {\bibinfo {volume} {133}},\ \bibinfo {pages} {086301} (\bibinfo {year}
  {2024})}\BibitemShut {NoStop}%
\bibitem [{\citenamefont {Pino}\ \emph {et~al.}(2025)\citenamefont {Pino},
  \citenamefont {Meir},\ and\ \citenamefont {Aguado}}]{pino2024thermodynamics}%
  \BibitemOpen
  \bibfield  {author} {\bibinfo {author} {\bibfnamefont {D.~M.}\ \bibnamefont
  {Pino}}, \bibinfo {author} {\bibfnamefont {Y.}~\bibnamefont {Meir}},\ and\
  \bibinfo {author} {\bibfnamefont {R.}~\bibnamefont {Aguado}},\ }\bibfield
  {title} {\bibinfo {title} {Thermodynamics of non-{H}ermitian {J}osephson
  junctions with exceptional points},\ }\href
  {https://doi.org/10.1103/PhysRevB.111.L140503} {\bibfield  {journal}
  {\bibinfo  {journal} {Phys. Rev. B}\ }\textbf {\bibinfo {volume} {111}},\
  \bibinfo {pages} {L140503} (\bibinfo {year} {2025})}\BibitemShut {NoStop}%
\bibitem [{\citenamefont {Ohnmacht}\ \emph {et~al.}(2025)\citenamefont
  {Ohnmacht}, \citenamefont {Wilhelm}, \citenamefont {Weisbrich},\ and\
  \citenamefont {Belzig}}]{Ohnmacht2024}%
  \BibitemOpen
  \bibfield  {author} {\bibinfo {author} {\bibfnamefont {D.~C.}\ \bibnamefont
  {Ohnmacht}}, \bibinfo {author} {\bibfnamefont {V.}~\bibnamefont {Wilhelm}},
  \bibinfo {author} {\bibfnamefont {H.}~\bibnamefont {Weisbrich}},\ and\
  \bibinfo {author} {\bibfnamefont {W.}~\bibnamefont {Belzig}},\ }\bibfield
  {title} {\bibinfo {title} {Non-{H}ermitian topology in multiterminal
  superconducting junctions},\ }\href
  {https://doi.org/10.1103/PhysRevLett.134.156601} {\bibfield  {journal}
  {\bibinfo  {journal} {Phys. Rev. Lett.}\ }\textbf {\bibinfo {volume} {134}},\
  \bibinfo {pages} {156601} (\bibinfo {year} {2025})}\BibitemShut {NoStop}%
\bibitem [{\citenamefont {Cayao}\ and\ \citenamefont
  {Sato}(2024{\natexlab{b}})}]{cayao2024non}%
  \BibitemOpen
  \bibfield  {author} {\bibinfo {author} {\bibfnamefont {J.}~\bibnamefont
  {Cayao}}\ and\ \bibinfo {author} {\bibfnamefont {M.}~\bibnamefont {Sato}},\
  }\bibfield  {title} {\bibinfo {title} {Non-{H}ermitian multiterminal
  phase-biased {J}osephson junctions},\ }\href
  {https://doi.org/10.1103/PhysRevB.110.235426} {\bibfield  {journal} {\bibinfo
   {journal} {Phys. Rev. B}\ }\textbf {\bibinfo {volume} {110}},\ \bibinfo
  {pages} {235426} (\bibinfo {year} {2024}{\natexlab{b}})}\BibitemShut
  {NoStop}%
\bibitem [{\citenamefont {Li}\ and\ \citenamefont
  {Trauzettel}(2025)}]{li2025EP}%
  \BibitemOpen
  \bibfield  {author} {\bibinfo {author} {\bibfnamefont {C.-A.}\ \bibnamefont
  {Li}}\ and\ \bibinfo {author} {\bibfnamefont {B.}~\bibnamefont
  {Trauzettel}},\ }\bibfield  {title} {\bibinfo {title} {Exceptional {A}ndreev
  spectrum and supercurrent in $p$-wave non-{H}ermitian {J}osephson
  junctions},\ }\href {https://doi.org/10.1103/58vd-b181} {\bibfield  {journal}
  {\bibinfo  {journal} {Phys. Rev. B}\ }\textbf {\bibinfo {volume} {112}},\
  \bibinfo {pages} {184504} (\bibinfo {year} {2025})}\BibitemShut {NoStop}%
\bibitem [{\citenamefont {Capecelatro}\ \emph {et~al.}(2025)\citenamefont
  {Capecelatro}, \citenamefont {Marciani}, \citenamefont {Campagnano},\ and\
  \citenamefont {Lucignano}}]{PhysRevB.111.064517}%
  \BibitemOpen
  \bibfield  {author} {\bibinfo {author} {\bibfnamefont {R.}~\bibnamefont
  {Capecelatro}}, \bibinfo {author} {\bibfnamefont {M.}~\bibnamefont
  {Marciani}}, \bibinfo {author} {\bibfnamefont {G.}~\bibnamefont
  {Campagnano}},\ and\ \bibinfo {author} {\bibfnamefont {P.}~\bibnamefont
  {Lucignano}},\ }\bibfield  {title} {\bibinfo {title} {Andreev non-{H}ermitian
  {H}amiltonian for open {J}osephson junctions from {G}reen's functions},\
  }\href {https://doi.org/10.1103/PhysRevB.111.064517} {\bibfield  {journal}
  {\bibinfo  {journal} {Phys. Rev. B}\ }\textbf {\bibinfo {volume} {111}},\
  \bibinfo {pages} {064517} (\bibinfo {year} {2025})}\BibitemShut {NoStop}%
\bibitem [{\citenamefont {Ogino}\ and\ \citenamefont
  {Uchino}(2025)}]{ogino2025}%
  \BibitemOpen
  \bibfield  {author} {\bibinfo {author} {\bibfnamefont {R.}~\bibnamefont
  {Ogino}}\ and\ \bibinfo {author} {\bibfnamefont {S.}~\bibnamefont {Uchino}},\
  }\bibfield  {title} {\bibinfo {title} {Anomalous supercurrents in the
  presence of particle losses},\ }\href {https://arxiv.org/abs/2505.21085}
  {\bibfield  {journal} {\bibinfo  {journal} {arXiv:2505.21085}\ } (\bibinfo
  {year} {2025})}\BibitemShut {NoStop}%
\bibitem [{\citenamefont {Solow}\ and\ \citenamefont
  {Flensberg}(2025)}]{solow2025EP}%
  \BibitemOpen
  \bibfield  {author} {\bibinfo {author} {\bibfnamefont {O.}~\bibnamefont
  {Solow}}\ and\ \bibinfo {author} {\bibfnamefont {K.}~\bibnamefont
  {Flensberg}},\ }\bibfield  {title} {\bibinfo {title} {Signatures of
  exceptional points in multiterminal superconductor--normal metal junctions},\
  }\href {https://doi.org/10.1103/dmfm-71l6} {\bibfield  {journal} {\bibinfo
  {journal} {Phys. Rev. B}\ }\textbf {\bibinfo {volume} {112}},\ \bibinfo
  {pages} {L161402} (\bibinfo {year} {2025})}\BibitemShut {NoStop}%
\bibitem [{\citenamefont {Qi}\ \emph {et~al.}(2025)\citenamefont {Qi},
  \citenamefont {Lu}, \citenamefont {Liu}, \citenamefont {Chen},\ and\
  \citenamefont {Xie}}]{JunjieNHDiode}%
  \BibitemOpen
  \bibfield  {author} {\bibinfo {author} {\bibfnamefont {J.}~\bibnamefont
  {Qi}}, \bibinfo {author} {\bibfnamefont {M.}~\bibnamefont {Lu}}, \bibinfo
  {author} {\bibfnamefont {J.}~\bibnamefont {Liu}}, \bibinfo {author}
  {\bibfnamefont {C.-Z.}\ \bibnamefont {Chen}},\ and\ \bibinfo {author}
  {\bibfnamefont {X.~C.}\ \bibnamefont {Xie}},\ }\bibfield  {title} {\bibinfo
  {title} {Non-{H}ermitian superconducting diode effect},\ }\href
  {https://doi.org/10.1103/n51c-17pn} {\bibfield  {journal} {\bibinfo
  {journal} {Phys. Rev. B}\ }\textbf {\bibinfo {volume} {112}},\ \bibinfo
  {pages} {L060502} (\bibinfo {year} {2025})}\BibitemShut {NoStop}%
\bibitem [{\citenamefont {Cayao}\ and\ \citenamefont
  {Sato}(2026{\natexlab{b}})}]{cayaoSatoNH4MZMs}%
  \BibitemOpen
  \bibfield  {author} {\bibinfo {author} {\bibfnamefont {J.}~\bibnamefont
  {Cayao}}\ and\ \bibinfo {author} {\bibfnamefont {M.}~\bibnamefont {Sato}},\
  }\bibfield  {title} {\bibinfo {title} {Non-{H}ermitian {J}osephson junctions
  with four {M}ajorana zero modes},\ }\href
  {https://doi.org/10.7566/JPSJ.95.014705} {\bibfield  {journal} {\bibinfo
  {journal} {J. Phys. Soc. Jpn.}\ }\textbf {\bibinfo {volume} {95}},\ \bibinfo
  {pages} {014705} (\bibinfo {year} {2026}{\natexlab{b}})}\BibitemShut
  {NoStop}%
\bibitem [{\citenamefont {Pay\'a}\ \emph {et~al.}(2026)\citenamefont {Pay\'a},
  \citenamefont {Solow}, \citenamefont {Prada}, \citenamefont {Aguado},\ and\
  \citenamefont {Flensberg}}]{9jdy-b418}%
  \BibitemOpen
  \bibfield  {author} {\bibinfo {author} {\bibfnamefont {C.}~\bibnamefont
  {Pay\'a}}, \bibinfo {author} {\bibfnamefont {O.}~\bibnamefont {Solow}},
  \bibinfo {author} {\bibfnamefont {E.}~\bibnamefont {Prada}}, \bibinfo
  {author} {\bibfnamefont {R.}~\bibnamefont {Aguado}},\ and\ \bibinfo {author}
  {\bibfnamefont {K.}~\bibnamefont {Flensberg}},\ }\bibfield  {title} {\bibinfo
  {title} {Non-{H}ermitian skin effect and electronic nonlocal transport},\
  }\href {https://doi.org/10.1103/9jdy-b418} {\bibfield  {journal} {\bibinfo
  {journal} {Phys. Rev. B}\ }\textbf {\bibinfo {volume} {113}},\ \bibinfo
  {pages} {L161405} (\bibinfo {year} {2026})}\BibitemShut {NoStop}%
\bibitem [{\citenamefont {Raj}\ \emph {et~al.}(2026)\citenamefont {Raj},
  \citenamefont {Ray},\ and\ \citenamefont {Samal}}]{mhwj-5c85}%
  \BibitemOpen
  \bibfield  {author} {\bibinfo {author} {\bibfnamefont {A.}~\bibnamefont
  {Raj}}, \bibinfo {author} {\bibfnamefont {S.}~\bibnamefont {Ray}},\ and\
  \bibinfo {author} {\bibfnamefont {S.~S.}\ \bibnamefont {Samal}},\ }\bibfield
  {title} {\bibinfo {title} {Symmetry and topology in a non-{H}ermitian
  {K}itaev chain},\ }\href {https://doi.org/10.1103/mhwj-5c85} {\bibfield
  {journal} {\bibinfo  {journal} {Phys. Rev. B}\ }\textbf {\bibinfo {volume}
  {113}},\ \bibinfo {pages} {235104} (\bibinfo {year} {2026})}\BibitemShut
  {NoStop}%
\bibitem [{\citenamefont {Capecelatro}\ \emph
  {et~al.}(2026{\natexlab{a}})\citenamefont {Capecelatro}, \citenamefont
  {Marciani}, \citenamefont {Campagnano}, \citenamefont {Citro},\ and\
  \citenamefont {Lucignano}}]{gbhj-65b7}%
  \BibitemOpen
  \bibfield  {author} {\bibinfo {author} {\bibfnamefont {R.}~\bibnamefont
  {Capecelatro}}, \bibinfo {author} {\bibfnamefont {M.}~\bibnamefont
  {Marciani}}, \bibinfo {author} {\bibfnamefont {G.}~\bibnamefont
  {Campagnano}}, \bibinfo {author} {\bibfnamefont {R.}~\bibnamefont {Citro}},\
  and\ \bibinfo {author} {\bibfnamefont {P.}~\bibnamefont {Lucignano}},\
  }\bibfield  {title} {\bibinfo {title} {Supercurrent from the imaginary part
  of the {A}ndreev levels in non-{H}ermitian {J}osephson junctions},\ }\href
  {https://doi.org/10.1103/gbhj-65b7} {\bibfield  {journal} {\bibinfo
  {journal} {Phys. Rev. B}\ }\textbf {\bibinfo {volume} {113}},\ \bibinfo
  {pages} {184506} (\bibinfo {year} {2026}{\natexlab{a}})}\BibitemShut
  {NoStop}%
\bibitem [{\citenamefont {Capecelatro}\ \emph
  {et~al.}(2026{\natexlab{b}})\citenamefont {Capecelatro}, \citenamefont
  {Marciani}, \citenamefont {Guarcello}, \citenamefont {Campagnano},
  \citenamefont {Lucignano},\ and\ \citenamefont {Citro}}]{capecelatro20260pi}%
  \BibitemOpen
  \bibfield  {author} {\bibinfo {author} {\bibfnamefont {R.}~\bibnamefont
  {Capecelatro}}, \bibinfo {author} {\bibfnamefont {M.}~\bibnamefont
  {Marciani}}, \bibinfo {author} {\bibfnamefont {C.}~\bibnamefont {Guarcello}},
  \bibinfo {author} {\bibfnamefont {G.}~\bibnamefont {Campagnano}}, \bibinfo
  {author} {\bibfnamefont {P.}~\bibnamefont {Lucignano}},\ and\ \bibinfo
  {author} {\bibfnamefont {R.}~\bibnamefont {Citro}},\ }\bibfield  {title}
  {\bibinfo {title} {$0-\pi$ transitions in non-{H}ermitian magnetic
  {J}osephson junctions},\ }\href {https://arxiv.org/abs/2604.17978} {\bibfield
   {journal} {\bibinfo  {journal} {arXiv: 2604.17978}\ } (\bibinfo {year}
  {2026}{\natexlab{b}})}\BibitemShut {NoStop}%
\bibitem [{\citenamefont {Jana}\ \emph {et~al.}(2021)\citenamefont {Jana},
  \citenamefont {Chowdhury},\ and\ \citenamefont {Saha}}]{PhysRevB.103.235438}%
  \BibitemOpen
  \bibfield  {author} {\bibinfo {author} {\bibfnamefont {S.}~\bibnamefont
  {Jana}}, \bibinfo {author} {\bibfnamefont {D.}~\bibnamefont {Chowdhury}},\
  and\ \bibinfo {author} {\bibfnamefont {A.}~\bibnamefont {Saha}},\ }\bibfield
  {title} {\bibinfo {title} {Emergence of exceptional points and their
  spectroscopic signature in a dirac semimetal--dirty superconductor
  heterojunction},\ }\href {https://doi.org/10.1103/PhysRevB.103.235438}
  {\bibfield  {journal} {\bibinfo  {journal} {Phys. Rev. B}\ }\textbf {\bibinfo
  {volume} {103}},\ \bibinfo {pages} {235438} (\bibinfo {year}
  {2021})}\BibitemShut {NoStop}%
\bibitem [{\citenamefont {Cayao}\ and\ \citenamefont
  {Aguado}(2025)}]{cayaominimalkitaev}%
  \BibitemOpen
  \bibfield  {author} {\bibinfo {author} {\bibfnamefont {J.}~\bibnamefont
  {Cayao}}\ and\ \bibinfo {author} {\bibfnamefont {R.}~\bibnamefont {Aguado}},\
  }\bibfield  {title} {\bibinfo {title} {Non-{H}ermitian minimal {K}itaev
  chains},\ }\href {https://doi.org/10.1103/PhysRevB.111.205432} {\bibfield
  {journal} {\bibinfo  {journal} {Phys. Rev. B}\ }\textbf {\bibinfo {volume}
  {111}},\ \bibinfo {pages} {205432} (\bibinfo {year} {2025})}\BibitemShut
  {NoStop}%
\bibitem [{\citenamefont {Ezawa}(2024)}]{PhysRevB.109.L161404}%
  \BibitemOpen
  \bibfield  {author} {\bibinfo {author} {\bibfnamefont {M.}~\bibnamefont
  {Ezawa}},\ }\bibfield  {title} {\bibinfo {title} {Even-odd effect on
  robustness of majorana edge states in short kitaev chains},\ }\href
  {https://doi.org/10.1103/PhysRevB.109.L161404} {\bibfield  {journal}
  {\bibinfo  {journal} {Phys. Rev. B}\ }\textbf {\bibinfo {volume} {109}},\
  \bibinfo {pages} {L161404} (\bibinfo {year} {2024})}\BibitemShut {NoStop}%
\bibitem [{\citenamefont {Bordin}\ \emph {et~al.}(2023)\citenamefont {Bordin},
  \citenamefont {Wang}, \citenamefont {Liu}, \citenamefont {ten Haaf},
  \citenamefont {van Loo}, \citenamefont {Mazur}, \citenamefont {Xu},
  \citenamefont {van Driel}, \citenamefont {Zatelli}, \citenamefont
  {Gazibegovic}, \citenamefont {Badawy}, \citenamefont {Bakkers}, \citenamefont
  {Wimmer}, \citenamefont {Kouwenhoven},\ and\ \citenamefont
  {Dvir}}]{PhysRevX.13.031031}%
  \BibitemOpen
  \bibfield  {author} {\bibinfo {author} {\bibfnamefont {A.}~\bibnamefont
  {Bordin}}, \bibinfo {author} {\bibfnamefont {G.}~\bibnamefont {Wang}},
  \bibinfo {author} {\bibfnamefont {C.-X.}\ \bibnamefont {Liu}}, \bibinfo
  {author} {\bibfnamefont {S.~L.~D.}\ \bibnamefont {ten Haaf}}, \bibinfo
  {author} {\bibfnamefont {N.}~\bibnamefont {van Loo}}, \bibinfo {author}
  {\bibfnamefont {G.~P.}\ \bibnamefont {Mazur}}, \bibinfo {author}
  {\bibfnamefont {D.}~\bibnamefont {Xu}}, \bibinfo {author} {\bibfnamefont
  {D.}~\bibnamefont {van Driel}}, \bibinfo {author} {\bibfnamefont
  {F.}~\bibnamefont {Zatelli}}, \bibinfo {author} {\bibfnamefont
  {S.}~\bibnamefont {Gazibegovic}}, \bibinfo {author} {\bibfnamefont
  {G.}~\bibnamefont {Badawy}}, \bibinfo {author} {\bibfnamefont {E.~P. A.~M.}\
  \bibnamefont {Bakkers}}, \bibinfo {author} {\bibfnamefont {M.}~\bibnamefont
  {Wimmer}}, \bibinfo {author} {\bibfnamefont {L.~P.}\ \bibnamefont
  {Kouwenhoven}},\ and\ \bibinfo {author} {\bibfnamefont {T.}~\bibnamefont
  {Dvir}},\ }\bibfield  {title} {\bibinfo {title} {Tunable crossed andreev
  reflection and elastic cotunneling in hybrid nanowires},\ }\href
  {https://doi.org/10.1103/PhysRevX.13.031031} {\bibfield  {journal} {\bibinfo
  {journal} {Phys. Rev. X}\ }\textbf {\bibinfo {volume} {13}},\ \bibinfo
  {pages} {031031} (\bibinfo {year} {2023})}\BibitemShut {NoStop}%
\bibitem [{\citenamefont {Bordin}\ \emph {et~al.}(2024)\citenamefont {Bordin},
  \citenamefont {Li}, \citenamefont {van Driel}, \citenamefont {Wolff},
  \citenamefont {Wang}, \citenamefont {ten Haaf}, \citenamefont {Wang},
  \citenamefont {van Loo}, \citenamefont {Kouwenhoven},\ and\ \citenamefont
  {Dvir}}]{bordin2023crossed}%
  \BibitemOpen
  \bibfield  {author} {\bibinfo {author} {\bibfnamefont {A.}~\bibnamefont
  {Bordin}}, \bibinfo {author} {\bibfnamefont {X.}~\bibnamefont {Li}}, \bibinfo
  {author} {\bibfnamefont {D.}~\bibnamefont {van Driel}}, \bibinfo {author}
  {\bibfnamefont {J.~C.}\ \bibnamefont {Wolff}}, \bibinfo {author}
  {\bibfnamefont {Q.}~\bibnamefont {Wang}}, \bibinfo {author} {\bibfnamefont
  {S.~L.~D.}\ \bibnamefont {ten Haaf}}, \bibinfo {author} {\bibfnamefont
  {G.}~\bibnamefont {Wang}}, \bibinfo {author} {\bibfnamefont {N.}~\bibnamefont
  {van Loo}}, \bibinfo {author} {\bibfnamefont {L.~P.}\ \bibnamefont
  {Kouwenhoven}},\ and\ \bibinfo {author} {\bibfnamefont {T.}~\bibnamefont
  {Dvir}},\ }\bibfield  {title} {\bibinfo {title} {Crossed andreev reflection
  and elastic cotunneling in three quantum dots coupled by superconductors},\
  }\href {https://doi.org/10.1103/PhysRevLett.132.056602} {\bibfield  {journal}
  {\bibinfo  {journal} {Phys. Rev. Lett.}\ }\textbf {\bibinfo {volume} {132}},\
  \bibinfo {pages} {056602} (\bibinfo {year} {2024})}\BibitemShut {NoStop}%
\bibitem [{\citenamefont {Kawabata}\ \emph
  {et~al.}(2019{\natexlab{b}})\citenamefont {Kawabata}, \citenamefont
  {Bessho},\ and\ \citenamefont {Sato}}]{PhysRevLett.123.066405}%
  \BibitemOpen
  \bibfield  {author} {\bibinfo {author} {\bibfnamefont {K.}~\bibnamefont
  {Kawabata}}, \bibinfo {author} {\bibfnamefont {T.}~\bibnamefont {Bessho}},\
  and\ \bibinfo {author} {\bibfnamefont {M.}~\bibnamefont {Sato}},\ }\bibfield
  {title} {\bibinfo {title} {Classification of exceptional points and
  non-{H}ermitian topological semimetals},\ }\href
  {https://doi.org/10.1103/PhysRevLett.123.066405} {\bibfield  {journal}
  {\bibinfo  {journal} {Phys. Rev. Lett.}\ }\textbf {\bibinfo {volume} {123}},\
  \bibinfo {pages} {066405} (\bibinfo {year} {2019}{\natexlab{b}})}\BibitemShut
  {NoStop}%
\bibitem [{\citenamefont {Mahan}(2013)}]{mahan2013many}%
  \BibitemOpen
  \bibfield  {author} {\bibinfo {author} {\bibfnamefont {G.~D.}\ \bibnamefont
  {Mahan}},\ }\href@noop {} {\emph {\bibinfo {title} {Many-particle physics}}}\
  (\bibinfo  {publisher} {Springer Science \& Business Media},\ \bibinfo {year}
  {2013})\BibitemShut {NoStop}%
\bibitem [{\citenamefont {Zagoskin}(2014)}]{zagoskin}%
  \BibitemOpen
  \bibfield  {author} {\bibinfo {author} {\bibfnamefont {A.}~\bibnamefont
  {Zagoskin}},\ }\href@noop {} {\emph {\bibinfo {title} {Quantum Theory of
  Many-Body Systems: Techniques and Applications}}}\ (\bibinfo  {publisher}
  {Springer},\ \bibinfo {year} {2014})\BibitemShut {NoStop}%
\bibitem [{\citenamefont {San-Jose}\ \emph {et~al.}(2012)\citenamefont
  {San-Jose}, \citenamefont {Prada},\ and\ \citenamefont
  {Aguado}}]{PhysRevLett.108.257001}%
  \BibitemOpen
  \bibfield  {author} {\bibinfo {author} {\bibfnamefont {P.}~\bibnamefont
  {San-Jose}}, \bibinfo {author} {\bibfnamefont {E.}~\bibnamefont {Prada}},\
  and\ \bibinfo {author} {\bibfnamefont {R.}~\bibnamefont {Aguado}},\
  }\bibfield  {title} {\bibinfo {title} {ac {J}osephson effect in finite-length
  nanowire junctions with {M}ajorana modes},\ }\href
  {https://doi.org/10.1103/PhysRevLett.108.257001} {\bibfield  {journal}
  {\bibinfo  {journal} {Phys. Rev. Lett.}\ }\textbf {\bibinfo {volume} {108}},\
  \bibinfo {pages} {257001} (\bibinfo {year} {2012})}\BibitemShut {NoStop}%
\bibitem [{\citenamefont {Pikulin}\ and\ \citenamefont
  {Nazarov}(2012)}]{PhysRevB.86.140504}%
  \BibitemOpen
  \bibfield  {author} {\bibinfo {author} {\bibfnamefont {D.~I.}\ \bibnamefont
  {Pikulin}}\ and\ \bibinfo {author} {\bibfnamefont {Y.~V.}\ \bibnamefont
  {Nazarov}},\ }\bibfield  {title} {\bibinfo {title} {Phenomenology and
  dynamics of a {M}ajorana {J}osephson junction},\ }\href
  {https://doi.org/10.1103/PhysRevB.86.140504} {\bibfield  {journal} {\bibinfo
  {journal} {Phys. Rev. B}\ }\textbf {\bibinfo {volume} {86}},\ \bibinfo
  {pages} {140504} (\bibinfo {year} {2012})}\BibitemShut {NoStop}%
\bibitem [{\citenamefont {Cayao}\ \emph {et~al.}(2015)\citenamefont {Cayao},
  \citenamefont {Prada}, \citenamefont {San-Jose},\ and\ \citenamefont
  {Aguado}}]{PhysRevB.91.024514}%
  \BibitemOpen
  \bibfield  {author} {\bibinfo {author} {\bibfnamefont {J.}~\bibnamefont
  {Cayao}}, \bibinfo {author} {\bibfnamefont {E.}~\bibnamefont {Prada}},
  \bibinfo {author} {\bibfnamefont {P.}~\bibnamefont {San-Jose}},\ and\
  \bibinfo {author} {\bibfnamefont {R.}~\bibnamefont {Aguado}},\ }\bibfield
  {title} {\bibinfo {title} {{SNS} junctions in nanowires with spin-orbit
  coupling: Role of confinement and helicity on the subgap spectrum},\ }\href
  {https://doi.org/10.1103/PhysRevB.91.024514} {\bibfield  {journal} {\bibinfo
  {journal} {Phys. Rev. B}\ }\textbf {\bibinfo {volume} {91}},\ \bibinfo
  {pages} {024514} (\bibinfo {year} {2015})}\BibitemShut {NoStop}%
\bibitem [{\citenamefont {Valentini}\ \emph {et~al.}(2014)\citenamefont
  {Valentini}, \citenamefont {Fazio},\ and\ \citenamefont
  {Taddei}}]{PhysRevB.89.014509}%
  \BibitemOpen
  \bibfield  {author} {\bibinfo {author} {\bibfnamefont {S.}~\bibnamefont
  {Valentini}}, \bibinfo {author} {\bibfnamefont {R.}~\bibnamefont {Fazio}},\
  and\ \bibinfo {author} {\bibfnamefont {F.}~\bibnamefont {Taddei}},\
  }\bibfield  {title} {\bibinfo {title} {Andreev levels spectroscopy of
  topological three-terminal junctions},\ }\href
  {https://doi.org/10.1103/PhysRevB.89.014509} {\bibfield  {journal} {\bibinfo
  {journal} {Phys. Rev. B}\ }\textbf {\bibinfo {volume} {89}},\ \bibinfo
  {pages} {014509} (\bibinfo {year} {2014})}\BibitemShut {NoStop}%
\bibitem [{\citenamefont {Cayao}\ \emph {et~al.}(2017)\citenamefont {Cayao},
  \citenamefont {San-Jose}, \citenamefont {Black-Schaffer}, \citenamefont
  {Aguado},\ and\ \citenamefont {Prada}}]{PhysRevB.96.205425}%
  \BibitemOpen
  \bibfield  {author} {\bibinfo {author} {\bibfnamefont {J.}~\bibnamefont
  {Cayao}}, \bibinfo {author} {\bibfnamefont {P.}~\bibnamefont {San-Jose}},
  \bibinfo {author} {\bibfnamefont {A.~M.}\ \bibnamefont {Black-Schaffer}},
  \bibinfo {author} {\bibfnamefont {R.}~\bibnamefont {Aguado}},\ and\ \bibinfo
  {author} {\bibfnamefont {E.}~\bibnamefont {Prada}},\ }\bibfield  {title}
  {\bibinfo {title} {Majorana splitting from critical currents in {J}osephson
  junctions},\ }\href {https://doi.org/10.1103/PhysRevB.96.205425} {\bibfield
  {journal} {\bibinfo  {journal} {Phys. Rev. B}\ }\textbf {\bibinfo {volume}
  {96}},\ \bibinfo {pages} {205425} (\bibinfo {year} {2017})}\BibitemShut
  {NoStop}%
\bibitem [{\citenamefont {Murani}\ \emph {et~al.}(2017)\citenamefont {Murani},
  \citenamefont {Chepelianskii}, \citenamefont {Gu\'eron},\ and\ \citenamefont
  {Bouchiat}}]{PhysRevB.96.165415}%
  \BibitemOpen
  \bibfield  {author} {\bibinfo {author} {\bibfnamefont {A.}~\bibnamefont
  {Murani}}, \bibinfo {author} {\bibfnamefont {A.}~\bibnamefont
  {Chepelianskii}}, \bibinfo {author} {\bibfnamefont {S.}~\bibnamefont
  {Gu\'eron}},\ and\ \bibinfo {author} {\bibfnamefont {H.}~\bibnamefont
  {Bouchiat}},\ }\bibfield  {title} {\bibinfo {title} {Andreev spectrum with
  high spin-orbit interactions: Revealing spin splitting and topologically
  protected crossings},\ }\href {https://doi.org/10.1103/PhysRevB.96.165415}
  {\bibfield  {journal} {\bibinfo  {journal} {Phys. Rev. B}\ }\textbf {\bibinfo
  {volume} {96}},\ \bibinfo {pages} {165415} (\bibinfo {year}
  {2017})}\BibitemShut {NoStop}%
\bibitem [{\citenamefont {Cayao}\ \emph {et~al.}(2018)\citenamefont {Cayao},
  \citenamefont {Black-Schaffer}, \citenamefont {Prada},\ and\ \citenamefont
  {Aguado}}]{cayao2018andreev}%
  \BibitemOpen
  \bibfield  {author} {\bibinfo {author} {\bibfnamefont {J.}~\bibnamefont
  {Cayao}}, \bibinfo {author} {\bibfnamefont {A.~M.}\ \bibnamefont
  {Black-Schaffer}}, \bibinfo {author} {\bibfnamefont {E.}~\bibnamefont
  {Prada}},\ and\ \bibinfo {author} {\bibfnamefont {R.}~\bibnamefont
  {Aguado}},\ }\bibfield  {title} {\bibinfo {title} {Andreev spectrum and
  supercurrents in nanowire-based {SNS} junctions containing {M}ajorana bound
  states},\ }\href {https://www.beilstein-journals.org/bjnano/articles/9/127}
  {\bibfield  {journal} {\bibinfo  {journal} {Beilstein J. Nanotechnol.}\
  }\textbf {\bibinfo {volume} {9}},\ \bibinfo {pages} {1339} (\bibinfo {year}
  {2018})}\BibitemShut {NoStop}%
\bibitem [{\citenamefont {Peng}\ \emph {et~al.}(2016)\citenamefont {Peng},
  \citenamefont {Pientka}, \citenamefont {Berg}, \citenamefont {Oreg},\ and\
  \citenamefont {von Oppen}}]{PhysRevB.94.085409}%
  \BibitemOpen
  \bibfield  {author} {\bibinfo {author} {\bibfnamefont {Y.}~\bibnamefont
  {Peng}}, \bibinfo {author} {\bibfnamefont {F.}~\bibnamefont {Pientka}},
  \bibinfo {author} {\bibfnamefont {E.}~\bibnamefont {Berg}}, \bibinfo {author}
  {\bibfnamefont {Y.}~\bibnamefont {Oreg}},\ and\ \bibinfo {author}
  {\bibfnamefont {F.}~\bibnamefont {von Oppen}},\ }\bibfield  {title} {\bibinfo
  {title} {Signatures of topological {J}osephson junctions},\ }\href
  {https://doi.org/10.1103/PhysRevB.94.085409} {\bibfield  {journal} {\bibinfo
  {journal} {Phys. Rev. B}\ }\textbf {\bibinfo {volume} {94}},\ \bibinfo
  {pages} {085409} (\bibinfo {year} {2016})}\BibitemShut {NoStop}%
\bibitem [{\citenamefont {Awoga}\ \emph {et~al.}(2019)\citenamefont {Awoga},
  \citenamefont {Cayao},\ and\ \citenamefont
  {Black-Schaffer}}]{PhysRevLett.123.117001}%
  \BibitemOpen
  \bibfield  {author} {\bibinfo {author} {\bibfnamefont {O.~A.}\ \bibnamefont
  {Awoga}}, \bibinfo {author} {\bibfnamefont {J.}~\bibnamefont {Cayao}},\ and\
  \bibinfo {author} {\bibfnamefont {A.~M.}\ \bibnamefont {Black-Schaffer}},\
  }\bibfield  {title} {\bibinfo {title} {Supercurrent detection of
  topologically trivial zero-energy states in nanowire junctions},\ }\href
  {https://doi.org/10.1103/PhysRevLett.123.117001} {\bibfield  {journal}
  {\bibinfo  {journal} {Phys. Rev. Lett.}\ }\textbf {\bibinfo {volume} {123}},\
  \bibinfo {pages} {117001} (\bibinfo {year} {2019})}\BibitemShut {NoStop}%
\bibitem [{\citenamefont {Cayao}\ and\ \citenamefont
  {Black-Schaffer}(2021)}]{PhysRevB.104.L020501}%
  \BibitemOpen
  \bibfield  {author} {\bibinfo {author} {\bibfnamefont {J.}~\bibnamefont
  {Cayao}}\ and\ \bibinfo {author} {\bibfnamefont {A.~M.}\ \bibnamefont
  {Black-Schaffer}},\ }\bibfield  {title} {\bibinfo {title} {Distinguishing
  trivial and topological zero-energy states in long nanowire junctions},\
  }\href {https://doi.org/10.1103/PhysRevB.104.L020501} {\bibfield  {journal}
  {\bibinfo  {journal} {Phys. Rev. B}\ }\textbf {\bibinfo {volume} {104}},\
  \bibinfo {pages} {L020501} (\bibinfo {year} {2021})}\BibitemShut {NoStop}%
\bibitem [{\citenamefont {Heiss}(2004)}]{heiss2004exceptional}%
  \BibitemOpen
  \bibfield  {author} {\bibinfo {author} {\bibfnamefont {W.}~\bibnamefont
  {Heiss}},\ }\bibfield  {title} {\bibinfo {title} {Exceptional points--their
  universal occurrence and their physical significance},\ }\href
  {https://link.springer.com/article/10.1023/B:CJOP.0000044002.05657.04}
  {\bibfield  {journal} {\bibinfo  {journal} {Czechoslov. J. Phys.}\ }\textbf
  {\bibinfo {volume} {54}},\ \bibinfo {pages} {1091} (\bibinfo {year}
  {2004})}\BibitemShut {NoStop}%
\bibitem [{\citenamefont {Heiss}(2012)}]{Heiss_2012}%
  \BibitemOpen
  \bibfield  {author} {\bibinfo {author} {\bibfnamefont {W.~D.}\ \bibnamefont
  {Heiss}},\ }\bibfield  {title} {\bibinfo {title} {The physics of exceptional
  points},\ }\href {https://doi.org/10.1088/1751-8113/45/44/444016} {\bibfield
  {journal} {\bibinfo  {journal} {J. Phys. A Math. Theor.}\ }\textbf {\bibinfo
  {volume} {45}},\ \bibinfo {pages} {444016} (\bibinfo {year}
  {2012})}\BibitemShut {NoStop}%
\end{thebibliography}%
\end{document}